\title{Generalized and Extended Product Codes}
\author{Mario Blaum and Steven Hetzler\\
IBM Almaden Research Center\\
San Jose, CA 95120\\
{\tt mmblaum,hetzler@us.ibm.com}
}
\date{}
\documentclass[12pt]{article}
\usepackage{latexsym}
\usepackage{multirow}
 \newtheorem{theo}{Theorem}[section]
 \newtheorem{lemma}{Lemma}[section]
 
 \newtheorem{defin}{Definition}[section]
 \newtheorem{ex}{Example}[section]
 
 \newtheorem{cor}{Corollary}[section]
 
\newtheorem{COROLLARY}{\indent Corollary}

\newtheorem{EXAMPLE}{\indent Example}

\newtheorem{THEOREM}{\indent Theorem}

\newtheorem{REMARK}{\indent Remark}

\newcommand{\ga}{\mbox{$\gamma$}}

\newcommand{\fullstop}{\hspace{-0.85em} {\bf .}}

\newcommand{\uv}{\mbox{$\underline{v}$}}

\newcommand{\uu}{\mbox{$\underline{u}$}}

\newcommand{\uc}{\mbox{$\underline{c}$}}

\newcommand{\hs}{\mbox{$\hat{s}$}}

\newcommand{\la}{\mbox{$\leftarrow$}}

\newcommand{\al}{\mbox{$\alpha$}}

\newcommand{\eq}{\mbox{$\, =\,$}}

\newcommand{\lan}{\mbox{$\langle$}}
\newcommand{\ran}{\mbox{$\rangle$}}

\newcommand{\qed}{\hfill$\Box$\\[1ex]}
\newcommand{\pf}{{\bf Proof: }}

\newcommand{\uw}{\mbox{$\underline{w}$}}

\newcommand{\xor}{\mbox{$\,\oplus\,$}}

\newcommand{\C}{\mbox{${\cal C}$}}

\newcommand{\cH}{\mbox{${\cal H}$}}

\newcommand{\cO}{\mbox{${\cal O}$}}

\newcommand{\br}{\\ }
\newcommand{\ce}{\begin{center}}
\newcommand{\cen}{\end{center}}

\newcommand{\ipb}{\begin{description}}
\newcommand{\ipn}{\end{description}}

\newcommand{\qb}{\begin{quote}}
\newcommand{\qn}{\end{quote}}

\newcommand{\tp}{\begin{titlepage}}
\newcommand{\tpn}{\end{titlepage}}
\newcommand{\zb}{\begin{figure}[hbtp]}
\newcommand{\zn}{\end{figure}}

\newcommand{\EQX}[1]{\begin{equation}\label{#1}}
\newcommand{\ENX}{\end{equation}}
\newcommand{\EQL}{\begin{eqnarray*}}
\newcommand{\ENL}{\end{eqnarray*}}
\newcommand{\EQLX}[1]{\begin{eqnarray}\label{#1}}
\newcommand{\ENLX}{\end{eqnarray}}

\newcommand{\open}{\begin{document}}
\newcommand{\close}{\end{document}}
\newcommand{\lfcr}[1]{\br\hspace*{#1em}}
\newenvironment{mat}[1]
{\left[\begin{array}{#1}}{\end{array}\right]}
\newcommand{\GAMMA}{\Gamma}
\newcommand{\DELTA}{\Delta}
\newcommand{\THETA}{\Theta}
\newcommand{\LAMBDA}{\Lambda}
\newcommand{\XI}{\Xi}
\newcommand{\PI}{\Pi}
\newcommand{\SIGMA}{\Sigma}
\newcommand{\UPSILON}{\Upsilon}
\newcommand{\PHI}{\Phi}
\newcommand{\PSI}{\Psi}
\newcommand{\OMEGA}{\Omega}
\newcommand{\bldgreek}[1]{\mbox{\boldmath $#1$}}
\newcommand{\bldbeta}{\bldgreek{\beta}}
\newcommand{\bldgamma}{\bldgreek{\gamma}}
\newcommand{\blddelta}{\bldgreek{\delta}}
\newcommand{\bldepsilon}{\bldgreek{\epsilon}}
\newcommand{\bldvarepsilon}{\bldgreek{\varepsilon}}
\newcommand{\bldzeta}{\bldgreek{\zeta}}
\newcommand{\bldeta}{\bldgreek{\eta}}
\newcommand{\bldtheta}{\bldgreek{\theta}}
\newcommand{\bldvartheta}{\bldgreek{\vartheta}}
\newcommand{\bldiota}{\bldgreek{\iota}}
\newcommand{\bldkappa}{\bldgreek{\kappa}}
\newcommand{\bldlambda}{\bldgreek{\lambda}}
\newcommand{\bldmu}{\bldgreek{\mu}}
\newcommand{\bldnu}{\bldgreek{\nu}}
\newcommand{\bldxi}{\bldgreek{\xi}}
\newcommand{\bldpi}{\bldgreek{\pi}}
\newcommand{\bldvarpi}{\bldgreek{\varpi}}
\newcommand{\bldrho}{\bldgreek{\rho}}
\newcommand{\bldvarrho}{\bldgreek{\varrho}}
\newcommand{\bldsigma}{\bldgreek{\sigma}}
\newcommand{\bldvarsigma}{\bldgreek{\varsigma}}
\newcommand{\bldtau}{\bldgreek{\tau}}
\newcommand{\bldupsilon}{\bldgreek{\upsilon}}
\newcommand{\bldphi}{\bldgreek{\phi}}
\newcommand{\bldvarphi}{\bldgreek{\varphi}}
\newcommand{\bldchi}{\bldgreek{\chi}}
\newcommand{\bldpsi}{\bldgreek{\psi}}
\newcommand{\bldomega}{\bldgreek{\omega}}
\begin{document}
\parindent=10pt
\maketitle
\begin{abstract}
Generalized Product (GPC) Codes, an unification of Product
Codes and Integrated Interleaved (II) Codes, are
presented.
Applications for approaches requiring local and
global parities are described. 
The more general problem of
extending product codes by adding global parities is studied and an
upper bound on the minimum distance of such codes is obtained.
Codes with one, two and three global parities whose minimum distances
meet the bound are presented.
Tradeoffs between optimality and field size are discussed.

\vspace{.3cm}

\end{abstract}

\section{Introduction}
\label{Introduction}

There has been considerable research lately on
codes with local and global properties for erasure correction (see
for instance~\cite{bhh}\cite{bh}\cite{bpsy}\cite{ghsy}\cite{hcl}\cite{kna}\cite{pd}\cite{pklk}\cite{rk}\cite{rp}\cite{sa}\cite{sd}\cite{tb}\cite{wz}
and references within).
In general, data symbols are divided into sets and parity symbols
(i.e., local parities) are added to each set (often, using an MDS code).
This way, when a number of erasures not exceeding the
number of parity symbols occurs in a set, such erasures are rapidly
recovered. In addition to the local parities, a number of
global parities are also added. Those global parities involve all of the
data symbols and may include the local parity symbols as well. The
global parities can correct situations in which the
erasure-correcting power of the local parities has been exceeded.

The interest in erasure correcting codes with local and global
properties arises mainly from
two applications. One of them is the cloud. A cloud configuration
may consist of many storage devices, of which some of them may even
be in different geographical locations and the data is distributed
across them. In the case that one or more of those devices fails, it is
desirable to recover its contents ``locally,'' that is, using a few
parity devices within a set of limited size in order to affect
performance as little as possible. However, the local parity may not
be enough. In case the erasure-correcting capability of a local set is
exceeded, extra protection is needed. In order to handle this
situation, some devices containing global parities are
incorporated, and when the local correction power is exceeded,
the global parities
are invoked and correction is attempted. If such a situation occurs,
there will be an impact on performance,
but data loss may be averted. It is expected that the cases in which
the local parity is
exceeded are relatively rare events, so the aforementioned impact on
performance does not occur frequently.
As an example of this type of application, we refer the reader to the
description of the Azure system~\cite{hsx} or to the Xorbas code
presented in~\cite{sa}.

A second application occurs in the context of Redundant Arrays of
Independent Disks (RAID) architectures~\cite{g}. In this case, a RAID
architecture protects against one or more storage device failures.
For example, RAID 5 adds one extra parity device, allowing for the
recovery of the contents of one failed device, while RAID 6 protects
against up to two device failures. In particular, if those devices
are Solid State Drives (SSDs), like flash memories, their
reliability decays with time and with the number of writes and
reads~\cite{M}.
The information in SSDs is generally divided into pages, each page
containing its own internal Error-Correction Code (ECC). It may
happen that a particular page degrades and its ECC is exceeded.
However, this situation may not be known to the user until the page is
accessed (what is known as a silent failure). Assuming an SSD has
failed in a RAID~5 scheme, if during reconstruction a silent page
failure is encountered in one of the surviving SSDs, then data loss
will occur. A method around this situation is using RAID~6. However,
this method is costly, since it requires two whole SSDs as parity. It is more
desirable to divide the information in a RAID type of architecture
into $m\times n$ stripes: $m$ represents the size of a stripe, and
$n$ is the number of SSDs. The RAID architecture can be viewed as
consisting of a large number of stripes, each stripe encoded and
decoded independently. Certainly, codes like the ones used in cloud
applications can be used as well for RAID applications. In practice,
the choice of code depends on the
statistics of errors and on the frequency of silent page failures.
RAID systems, however, may behave differently than a cloud array of devices, in
the sense that each column represents a whole storage device. When a device
fails, then the whole column is lost, a correlation that may not
occur in cloud applications. For that reason, RAID architectures may
benefit from a special class of codes with local and global
properties, the so called Sector-Disk (SD)
codes, which take into
account such correlations~\cite{pb}\cite{pbh}.

From now on, we call symbols the entries of a code with local and
global properties. Such symbols can be whole devices (for example, in
the case of cloud applications) or pages (in the case of RAID
applications for SSDs). Each symbol may be protected
by one local group, but a natural extension is to consider multiple
localities~\cite{rp}\cite{tb}\cite{zy}. A special case of multiple localities
is given by product codes~\cite{ms}: any symbol is protected by either
horizontal or vertical parities.

Product codes by themselves may also be used in RAID type of
architectures: the horizontal parities protect a number of devices
from failure. The vertical parities allow for rapid recovery of a page
or sector within a device (a first responder type of approach).
However, if the number of silent failures exceeds the correcting
capability of the vertical code, and the horizontal code is unusable
due to device failure, data loss will occur. For that reason, it may be
convenient to incorporate a number of global parities to the product
code.

In effect, assume that we have a product code consisting of $m\times
n$ arrays such that each column has $v$ parity symbols and each row
has $h$ parity symbols. If in addition to the horizontal and vertical
parities we have $g$ extra parities, we say that the code is an
Extended Product (EPC) code and we denote it by $EP(m,v;n,h;g)$.
Notice that if $g\eq 0$, we have a regular product code. Similarly,
if $v\eq 0$, we have a Locally Recoverable (LRC) code.

Constructions of LRC codes involve different issues and tradeoffs,
like the size of the field and optimality criteria. The same is true
for EPC codes, of which, as we have seen above, LRC codes are a
special case. In particular, one goal is to keep the size of the required
finite field small, since operations over a small field have
less complexity than over a larger field due to the smaller look-up
tables involved. For example, Integrated Interleaved
(II) codes~\cite{hapkt}\cite{tk}
over $GF(q)$, where $q\geq \max\{m,n\}$, were proposed in~\cite{bh}
as LRC codes (II codes are closely related to Generalized
Concatenated Codes~\cite{bz}\cite{z}). Let us mention the construction in~\cite{ll},
which also reduces field
size when failures are correlated. Similarly, we will propose a new
family of codes that
we call Generalized Product (GPC) codes, of which both product codes
and II codes are special cases.

As LRC codes, EPC codes also have optimality issues. For example,
LRC codes optimizing the minimum distance were presented
in~\cite{tb}, and except for special cases, in general II codes are
not optimal as LRC codes, but the codes in~\cite{tb} require a field
of size at least $mn$, so there is a tradeoff. The same happens with
GPC codes: except for the special case of one global parity, they do
not optimize the minimum
distance. We examine some cases of EPC codes that do optimize the
minimum distance for two and three global parities, but a larger field is
required.

There are stronger criteria for optimization than the minimum
distance in LRC codes. For example, PMDS
codes~\cite{bhh}\cite{bpsy}\cite{ghjy}\cite{hsx} satisfy the
Maximally Recoverable (MR) property\cite{ghjy}\cite{ghswy}. The
definition of the MR
property is extended for EPC codes in~\cite{ghswy}, but it turns out
that EPC codes with the MR property are difficult to obtain. For
example, in~\cite{ghswy} it was proven that an EPC code
$EP(m,1;n,1;1)$ (i.e., one vertical and one horizontal parity per column
and row and one global parity) with the MR property requires a field
that is superlinear on the size of the array (and no explicit
construction is given). We will not address EPC codes with the MR
property here.

Although the constructions can be extended to finite fields of any
characteristic, for simplicity, in what follows we assume that the
finite fields have characteristic 2.

The paper is structured as follows: in Section~\ref{GP} we present
the definition of GPC codes and give their properties, like their
erasure-correcting capability, their minimum distance and encoding and
decoding algorithms. In Section~\ref{optimality}, we present an upper
bound on the minimum distance of EPC codes and we give constructions
with one, two and three global parities attaining the bound. We end
the paper by drawing some conclusions.

\section{Generalized Product (GPC) Codes}
\label{GP}
We start by defining Generalized Product Codes, which unify
product codes with II codes. These codes also
consist of $m\times n$
arrays whose elements are in a finite field $GF(q)$ and it has
similar characteristics to a $t$-level II code, except that the last
$m-k$ rows are devoted to parity in such a way that each column in
the code belongs in an $[m,m-k]$ MDS code. Explicitly,

\begin{defin}
\label{defGPMDS}
{\rm
Take $t$ integers $1\leq u_0<u_1<\ldots <u_{t-1}\leq n-1$ and let
$\uu$ be the following vector of length
$m\eq s_0+s_1+\cdots +s_{t-1}$, where $s_i\geq 1$ for $0\leq i\leq t-1$:

\begin{eqnarray}
\label{equu}
\uu &=&
\left(\overbrace{u_0,u_0,\ldots,u_0}^{s_0},\overbrace{u_1,u_1,\ldots,u_1}^{s_1},\ldots,
\overbrace{u_{t-1},u_{t-1},\ldots,u_{t-1}}^{s_{t-1}}\right).
\end{eqnarray}

Consider a set $\{\C_i\}$ of $t$ nested $[n,n-u_i,u_i+1]$, $0\leq i\leq t-1$,
Reed-Solomon~\cite{ms} (RS) codes with elements in a finite field
$GF(q)$, $q\,>\,\max\{m,n\}$, such that
a parity-check matrix for $\C_i$ is given by

\begin{eqnarray}
\label{Hi}
H_i &=&
\left(
\begin{array}{ccccc}
1&1&1&\ldots &1\\
1&\al&\al^2&\ldots &\al^{n-1}\\
1&\al^2&\al^4&\ldots &\al^{2(n-1)}\\
\vdots &\vdots &\vdots &\ddots &\vdots \\
1&\al^{u_i-1}&\al^{2(u_i-1)}&\ldots &\al^{(u_i-1)(n-1)}\\
\end{array}
\right),
\end{eqnarray}
where $\al$ is an element of order $\cO(\al)\geq n$ in $GF(q)$.

For $0\leq m-k\,<\,s_{t-1}$, let $\C(n;k,\uu)$ be the code
consisting of $m\times n$ arrays over
$GF(q)$ such that, for each array in the code with rows
$\uc_0,\uc_1,\ldots,\uc_{m-1}$, $\uc_j\in\C_0$ for $0\leq
j\leq m-1$ and, if 

\begin{eqnarray}
\label{hsi}
\hs_i &=&
\sum_{j=i}^{t-1}s_j\quad {\rm for}\quad 0\leq i\leq t-1,
\end{eqnarray}
then
\begin{eqnarray}
\label{eqGP1}
\bigoplus_{j=0}^{m-1}\al^{rj}\uc_j&\in& \C_{i}\;\;{\rm for}\;\; 1\leq
i\leq t-1\;\; {\rm and}\;\; 0\leq r\leq \hs_{i}-1\\
\label{eqGP2}
\bigoplus_{j=0}^{m-1}\al^{rj}\uc_j&=& 0\;\;{\rm for}\;\; 0\leq
r\leq m-k-1.
\end{eqnarray}

Then we say that $\C(n;k,\uu)$ is a $t$-level Generalized Product (GPC) code.

\qed
}
\end{defin}

In reality, it is not necessary that the codes $\C_i$ in
Definition~\ref{defGPMDS} are RS with a parity-check matrix as given
by~(\ref{Hi}), or not even MDS, but we make the
assumption for simplicity. The codes may even be binary~\cite{w}.

Before giving the properties of $t$-level GPC codes, we present some examples.

\begin{ex}
\label{ex0}
{\em
Assume that $k\eq m$ in Definition~\ref{defGPMDS}, then, there are no
conditions~(\ref{eqGP2}) and
$\C(n;m,\uu)$ is a $t$-level Integrated
Interleaved (II)~\cite{bh}\cite{tk} code.

So, $t$-level II codes can be viewed as a special case of $t$-level GPC
codes.

\qed
}
\end{ex}

\begin{ex}
\label{ex4}
{\em
Assume that $t\eq 1$, then~(\ref{equu}) gives $\uu\eq
(\overbrace{u_0,u_0,\ldots,u_0}^{m})$ and, if $k<m$,\\
$\C(n;k,\overbrace{u_0,u_0,\ldots,u_0}^{m})$
is a regular product code~\cite{ms} such that each row
is in an $[n,n-u_0]$ code and each column in an $[m,k]$ code.

So, product codes can be viewed as a special case of $t$-level GPC codes.

\qed
}
\end{ex}

\begin{ex}
\label{ex5}
{\em
Assume that $t\eq 2$. Then, $\C_1\subset\C_0$, $\uu\eq
(\overbrace{u_0,u_0,\ldots,u_0}^{s_0},\overbrace{u_1,u_1,\ldots,u_1}^{s_1})$,\\
$s_0+s_1\eq m$,
and consider the 2-level GPC code $\C(n;k,\uu)$ with
$0\leq m-k < s_1$. Let $\uc\eq (\uc_0,\uc_1,\ldots,\uc_{m-1})$ be an
$m\times n$ array in $\C(n;k,\uu)$. Then, $\uc_j\in\C_0$ for each
$0\leq j\leq m-1$, and~(\ref{eqGP1}) and~(\ref{eqGP2}) give

\begin{eqnarray}
\label{eqGP1e2}
\bigoplus_{j=0}^{m-1}\al^{rj}\uc_j&\in& \C_{1}\;\;{\rm for}\;\;0\leq r\leq s_1-1\\
\label{eqGP2e2}
\bigoplus_{j=0}^{m-1}\al^{rj}\uc_j&=& 0\;\;{\rm for}\;\; 0\leq
r\leq m-k-1.
\end{eqnarray}

The 2-level II codes presented in~\cite{hapkt} correspond to
$\C(n;m,\uu)$ in this example, i.e., only equations (\ref{eqGP1e2})
are taken into account since $k\eq m$.

As another special case, take $k\eq m-1$ and
$\uu\eq (\overbrace{1,1,\ldots,1}^{m-2},2,2)$.
The rows $\uc_0,\uc_1,\ldots,\uc_{m-1}$ of
$\C(n;m-1,(\overbrace{1,1,\ldots,1}^{m-2},2,2))$
constitute a 2-level II code. Each column is in an $[m,m-1,2]$ code,
each row is in an $[n,n-1,2]$ code (single parity). The $\C_0$ code
is the $[n,n-1,2]$ code, and the $\C_1$ code is an $[n,n-2,3]$ code
given by the parity-check matrix

\begin{eqnarray*}
H_2 &=&
\left(
\begin{array}{ccccc}
1&1&1&\ldots &1\\
1&\al&\al^2&\ldots &\al^{n-1}\\
\end{array}
\right).
\end{eqnarray*}
Moreover, (\ref{eqGP1e2}) and~(\ref{eqGP2e2}) give

\begin{eqnarray*}
\bigoplus_{i=0}^{m-1}\al^i\uc_i&\in &\C_1\\
\bigoplus_{i=0}^{m-1}\uc_i&\eq &0.\\
\end{eqnarray*}

It is not hard to prove directly that this code can correct any 5
erasures, but this will be a consequence of Corollary~\ref{cor11}
to be presented below. It consists of a
product code (which has minimum distance 4) plus one extra (global) parity.
This extra parity brings the minimum distance up from 4 to 6.
For instance, if $m\eq 4$ and $n\eq 5,$
erasure patterns like the following (vertices of a rectangle)
$$
\begin{array}{|c|c|c|c|c|}
\hline
\phantom{X}&\phantom{X}&\phantom{X}&\phantom{X}&\phantom{X}\\
\hline
\phantom{X}&E&\phantom{X}&\phantom{X}&E\\
\hline
\phantom{X}&\phantom{X}&\phantom{X}&\phantom{X}&\phantom{X}\\
\hline
\phantom{X}&E&\phantom{X}&\phantom{X}&E\\
\hline
\end{array}
$$
are uncorrectable by the product code but not by
$\C(5;3,(1,1,2,2))$. An extra erasure in addition to the four
depicted above can be corrected by either the horizontal or the
vertical code.

\qed
}
\end{ex}

\begin{ex}
\label{ex6}
{\em
Assume that $t\eq 3$. Then, $\C_2\subset\C_1\subset\C_0$, $$\uu\eq
\left(\overbrace{u_0,u_0,\ldots,u_0}^{s_0},\overbrace{u_1,u_1,\ldots,u_1}^{s_1},
\overbrace{u_2,u_2,\ldots,u_2}^{s_2}\right),$$
$s_0+s_1+s_2\eq m$,
and consider the 3-level GPC code $\C(n;k,\uu)$ with
$0\leq m-k\leq s_2$. Let $\uc\eq (\uc_0,\uc_1,\ldots,\uc_{m-1})$ be an
$m\times n$ array in $\C(n;k,\uu)$. Then, $\uc_j\in\C_0$ for each
$0\leq j\leq m-1$, and~(\ref{eqGP1}) and~(\ref{eqGP2}) give

\begin{eqnarray}
\label{eqGP1e3}
\bigoplus_{j=0}^{m-1}\al^{rj}\uc_j&\in& \C_{2}\;\;{\rm for}\;\;0\leq
r\leq s_2-1\\
\label{eqGP1e3b}
\bigoplus_{j=0}^{m-1}\al^{rj}\uc_j&\in& \C_{1}\;\;{\rm for}\;\;0\leq r\leq s_1+s_2-1\\
\label{eqGP2e3}
\bigoplus_{j=0}^{m-1}\al^{rj}\uc_j&=& 0\;\;{\rm for}\;\; 0\leq
r\leq m-k-1.
\end{eqnarray}

\qed
}
\end{ex}

We are now ready to state the main result regarding GPC codes.

\begin{theo}
\label{theo2}
{\em
Consider an $m\times n$ array corresponding to a $\C(n;k,\uu)$
$t$-level GPC code as given by Definition~\ref{defGPMDS}. Then, the code
can correct up to $u_0$ erasures in any row, up
to $u_i$ erasures in any $s_i$ rows, $1\leq i\leq t-1$,
and up to $n$ erasures
in any $m-k$ rows.
}
\end{theo}

\noindent\pf
We may assume that the rows with erasures contain more than $u_0$
erasures,
since each row is in $\C_0$, an $[n,n-u_0,u_0+1]$ code, hence,
rows with up to $u_0$ erasures can be corrected.
Assume that there are up to $m-k$ erased rows and a number $\ell$ of rows
with more than $u_0$ erasures such that there are up
to $u_i$ erasures in any up to $s_i$ rows, $1\leq i\leq t-1$.
We do induction on $\ell$.

Assume first that $\ell\eq 0$, that is, we have up to $m-k$ erased
rows and the rest of the rows are erasure free. We can certainly
correct such up to $m-k$ erased rows by
using~(\ref{eqGP2}) (which states that each column in the array is in
an $[m,m-k,m-k+1]$ MDS code).

So, assume that there are $\ell\geq 1$ rows with more than $s_0$
erasures each
such that there are up
to $u_i$ erasures in any up to $s_i$ rows, $1\leq i\leq t-1$.
By induction, up to $\ell -1$ rows with this property are correctable.


Let
$i_0,i_1,\ldots,i_{m-1}$
be an ordering of the rows according to a
non-increasing number of erasures such that:

\begin{enumerate}

\item Rows $i_0,i_1,\ldots, i_{m-k-1}$
are erased.

\item Row $i_{m-k+j}$ for $0\leq
j\leq \ell-1$ has $v_j$ erasures, where $u_{t-1}\geq v_0\geq v_1\geq
\ldots\geq v_{\ell-1}\,>\,u_0$.

\item Rows $i_{m-k+\ell},i_{m-k+\ell+1},\ldots,i_{m-1}$ have no erasures.

\end{enumerate}

It suffices to prove that the $v_{\ell-1}$ erasures in row $i_{m-k+\ell-1}$ can be
corrected. Then we are left with $\ell -1$ rows with more than $s_0$
erasures each such that there are up
to $u_i$ erasures in any up to $s_i$ rows, $1\leq i\leq t-1$, 
and the result
follows by induction.

Choose a code $\C_s$ from the nested set of codes $\C_i$, $1\leq i\leq
t-1$, in Definition~\ref{defGPMDS} such that $\C_s$ can correct
$v_{\ell-1}$ erasures. Rearranging the order of the elements of the sums
in~(\ref{eqGP1}), and since
$\C_{t-1}\subset\C_{t-2}\subset\cdots\subset\C_s$, from~(\ref{eqGP1})
we have

\begin{eqnarray}
\label{eqGP11}
\bigoplus_{j=0}^{m-1}\al^{ri_j}\uc_{i_j}
&\in& \C_{s}\;\;{\rm for}\;\; 0\leq r\leq m-k+\ell -1.
\end{eqnarray}
Since the
$(m-k+\ell)\times m$
matrix corresponding to the
coefficients of the $\uc_{i_j}$s
in~(\ref{eqGP11}) is a Vandermonde matrix, it can be triangulated, giving

\begin{eqnarray}
\label{eqGP11t}
\uc_{i_r}\xor
\left(\bigoplus_{j=r+1}^{m-1}\ga_{r,j}\uc_{i_j}\right)
&\in& \C_{s}\;\;{\rm for}\;\; 0\leq r\leq m-k+\ell -1,
\end{eqnarray}
where the coefficients $\ga_{r,j}$ are a result of the triangulation.
In particular, taking\\ $r\eq m-k+\ell -1$ in~(\ref{eqGP11t}), we
obtain

\begin{eqnarray}
\label{eqGP11tl}
\uc_{i_{m-k+\ell -1}}
\xor
\left(\bigoplus_{j=m-k+\ell}^{m-1}\ga_{m-k+\ell -1,j}\uc_{i_j}\right)
&\in& \C_{s}.
\end{eqnarray}
Since $\uc_{i_{m-k+\ell -1}}$ has $v_{\ell-1}$ erasures and
$\uc_{i_j}$ has no erasures for $m-k+\ell\leq j\leq m-1$, then
$\uc_{i_{m-k+\ell -1}}
\xor
\left(\bigoplus_{j=m-k+\ell}^{m-1}\ga_{m-k+\ell
-1,j}\uc_{i_j}\right)$ has $v_{\ell-1}$ erasures. Since the vector is
in $\C_s$, the erasures can be corrected.
Once $\uc_{i_{m-k+\ell -1}}
\xor
\left(\bigoplus_{j=m-k+\ell}^{m-1}\ga_{m-k+\ell
-1,j}\uc_{i_j}\right)$ is corrected, $\uc_{i_{m-k+\ell -1}}$ is
obtained as

\begin{eqnarray*}
\uc_{i_{m-k+\ell -1}}&=&
\left(\uc_{i_{m-k+\ell -1}}\xor
\left(\bigoplus_{j=m-k+\ell}^{m-1}\ga_{m-k+\ell
-1,j}\uc_{i_j}\right)\right)\xor\left(\bigoplus_{j=m-k+\ell}^{m-1}\ga_{m-k+\ell
-1,j}\uc_{i_j}\right)
\end{eqnarray*}
and the result follows by induction on $\ell$.

\qed

Theorem~\ref{theo2} generalizes Theorem~1 in~\cite{bh}.
The proof of Theorem~\ref{theo2} is constructive in the sense that it
provides a decoding algorithm. The following example illustrates
Theorem~\ref{theo2} and the decoding algorithm.

\begin{ex}
\label{ex7}
{\em
Consider the 3-level GPC code $\C(7;4,(1,1,3,4,4,4))$ according to
Definition~\ref{defGPMDS} and Example~\ref{ex6}.
We have three codes $\C_2\subset\C_1\subset\C_0$, where
$\C_0$ is a $[7,6,2]$ code, $\C_1$ is a $[7,4,4]$ code and $\C_2$ is
a $[7,3,5]$ code. In addition, each column is in a $[6,4,3]$ code. We
may assume that the entries of these codes are in $GF(8)$ and that $\al$ is a
primitive element in $GF(8)$.

Consider the following $6\times 7$ array with erasures denoted by $E$:

$$
\begin{array}{rl}
\begin{array}{c}
\uc_0\\\uc_1\\\uc_2\\\uc_3\\\uc_4\\\uc_5\\
\end{array}
&
\begin{array}{|c|c|c|c|c|c|c|}
\hline
\phantom{X}&\phantom{X}&E&\phantom{X}&\phantom{X}&\phantom{X}&\phantom{X}\\
\hline
E&E&E&E&E&E&E\\
\hline
\phantom{X}&E&E&\phantom{X}&E&\phantom{X}&E\\
\hline
E&\phantom{X}&\phantom{X}&E&\phantom{X}&E&\phantom{X}\\
\hline
E&E&E&E&E&E&E\\
\hline
\phantom{X}&\phantom{X}&\phantom{X}&\phantom{X}&\phantom{X}&E&\phantom{X}\\
\hline
\end{array}
\end{array}
$$

The first step is correcting the single erasures in $\uc_0$ and in
$\uc_5$. An ordering of the remaining rows in non-increasing number
of erasures is $\{i_0,i_1,i_2,i_3\}\eq\{1,4,2,3\}$. In particular,
$\uc_3$ has three erasures. Following the proof of
Theorem~\ref{theo2}, there are $m-k\eq 2$ erased rows (rows $\uc_1$ and $\uc_4$)
and $\ell\eq 2$ rows with erasures, but not totally erased (rows $\uc_2$
and $\uc_3$). 
According to~(\ref{eqGP1e3}) and~(\ref{eqGP1e3b}),


$$
\begin{array}{ccccccccccccl}
\uc_0&\xor &\uc_1&\xor &\uc_2&\xor &\uc_3&\xor &\uc_4&\xor &\uc_5&\in &\C_2\\
\uc_0&\xor &\al \uc_1&\xor &\al^2\uc_2&\xor &\al^3\uc_3&\xor &\al^4\uc_4&\xor &\al^5\uc_5&\in
&\C_2\\
\uc_0&\xor &\al^2\uc_1&\xor &\al^4\uc_2&\xor &\al^6\uc_3&\xor &\al^8\uc_4&\xor &\al^{10}\uc_5&\in &\C_2\\
\uc_0&\xor &\al^3\uc_1&\xor &\al^6\uc_2&\xor &\al^9\uc_3&\xor &\al^{12}\uc_4&\xor &\al^{15}\uc_5&\in &\C_1.\\
\end{array}
$$

Notice that $\C_1$ can correct three erasures (i.e., $s\eq 1$ in the
proof of Theorem~\ref{theo2}).
Rearranging the $\uc_i$s above in non-increasing number
of erasures, we obtain


$$
\begin{array}{ccccccccccccl}
\uc_1&\xor &\uc_4&\xor &\uc_2&\xor &\uc_3&\xor &\uc_0&\xor &\uc_5&\in &\C_2\\
\al \uc_1&\xor &\al^4\uc_4&\xor &\al^2\uc_2&\xor &\al^3\uc_3&\xor &\uc_0&\xor &\al^5\uc_5&\in
&\C_2\\
\al^2\uc_1&\xor &\al^8\uc_4&\xor &\al^4\uc_2&\xor &\al^6\uc_3&\xor &\uc_0&\xor &\al^{10}\uc_5&\in &\C_2\\
\al^3\uc_1&\xor &\al^{12}\uc_4&\xor &\al^6\uc_2&\xor &\al^9\uc_3&\xor &\uc_0&\xor &\al^{15}\uc_5&\in &\C_1,\\
\end{array}
$$
which corresponds to~(\ref{eqGP11}) in the
proof of Theorem~\ref{theo2} (notice, $\C_2\subset\C_1$).
The coefficients in the linear system above correspond to the following
matrix:

\begin{eqnarray*}
\left(
\begin{array}{cccccc}
1&1&1&1&1&1\\
\al &\al^4 & \al^2&\al^3& 1 &\al^5\\
\al^2 &\al^8 & \al^4&\al^6& 1 &\al^{10}\\
\al^3 &\al^{12} & \al^6&\al^9& 1 &\al^{15}\\
\end{array}
\right).
\end{eqnarray*}

Triangulating this matrix in $GF(8)$, where $1\xor\al\xor\al^3\eq 0$,
gives

\begin{eqnarray*}
\left(
\begin{array}{cccccc}
1&1&1&1&1&1\\
0&1 & \al^2&\al^5& \al &\al^4\\
0&0&1&\al & \al^3 &\al\\
0&0&0&1& \al^3 &\al^{5}\\
\end{array}
\right).
\end{eqnarray*}

Applying this triangulation to the linear system, and since
$\C_2\subset\C_1$, we obtain the
following triangulated system:


$$
\begin{array}{ccccccccccccl}
\uc_1&\xor &\uc_4&\xor &\uc_2&\xor &\uc_3&\xor &\uc_0&\xor &\uc_5&\in &\C_2\\
&&\uc_4&\xor &\al^2\uc_2&\xor &\al^5\uc_3&\xor &\al \uc_0&\xor &\al^4\uc_5&\in
&\C_2\\
&&&&\uc_2&\xor &\al \uc_3&\xor &\al^3\uc_0&\xor &\al \uc_5&\in &\C_2\\
&&&&&&\uc_3&\xor &\al^3\uc_0&\xor &\al^{5}\uc_5&\in &\C_1.\\
\end{array}
$$

Since $\uc_3$ has 3 erasures and $\uc_0$ and $\uc_5$ have no
erasures, $\uc_3\xor\al^3\uc_0\xor\al^{5}\uc_5$ has 3
erasures, which can be corrected in $\C_1$.
Then, $$\uc_3\eq (\uc_3\xor\al^3\uc_0\xor\al^{5}\uc_5)\xor
(\al^3\uc_0\xor\al^{5}\uc_5).$$

Similarly, $\uc_2\xor\al \uc_3\xor\al^3\uc_0\xor\al \uc_5$ has 4
erasures, which can be corrected in $\C_2$,
and $$\uc_2\eq (\uc_2\xor\al \uc_3\xor\al^3\uc_0\xor\al \uc_5)\xor
(\al \uc_3\xor\al^3\uc_0\xor\al \uc_5).$$

Finally, $\uc_1$ and $\uc_4$ are obtained from~(\ref{eqGP2}).
We can apply the triangulation, so we obtain

\begin{eqnarray*}
\uc_4&\eq &\al^2\uc_2\xor\al^5\uc_3\xor\al \uc_0\xor\al^4\uc_5\\
\uc_1&\eq &\uc_4\xor\uc_2\xor\uc_3\xor\uc_0\xor\uc_5,
\end{eqnarray*}
completing the decoding.

\qed
}
\end{ex}

Before discussing the dimension, the encoding and the minimum
distance of the code, let us state and prove the following lemma.

\begin{lemma}
\label{lemma1}
{\em
Consider the $t$-level GPC code $\C(n;k,\uu)$ as given
by Definition~\ref{defGPMDS}. Then, if 
$\hs_t\eq m-k$ and $\hs_j$ is given by~(\ref{hsi})
for $0\leq j\leq t-1$, given $u_j+1$ fixed locations in $\hs_{j+1}+1$
different rows, then there is an array in $\C(n;k,\uu)$ that is
non-zero in such $\left(\hs_{j+1}+1\right)\left(u_j+1\right)$
locations and 0 elsewhere.
}
\end{lemma}

\noindent\pf
Given $j$ such that $0\leq j\leq t-1$ and $u_j+1$ fixed locations in a vector of length $n$, since
$\C_j$ is an $[n,n-u_j,u_j+1]$ MDS code, there is a codeword
$\uw$ in $\C_j$ whose non-zero entries are in such $u_j+1$ fixed
locations. Assume that the $\hs_{j+1}+1$ rows selected are
$i_0,i_1,\ldots,i_{\hs_{j+1}}$, where $$0\leq i_0\,<\,i_1 \,<\,\ldots
\,<\,i_{\hs_{j+1}}\leq m-1.$$
Let $\uv\,=\,(v_0,v_1,\ldots,v_{\hs_{j+1}})$ be a
codeword of weight $\hs_{j+1}+1$ in the (shortened) $[\hs_{j+1}+1,1,\hs_{j+1}+1]$ RS code whose
parity-check matrix is given by
$$
\left(
\begin{array}{ccccc}
1&1&1&\ldots &1\\
1&\al^{i_0} &\al^{i_1}& \ldots &\al^{i_{\hs_{j+1}}}\\
1&\al^{2i_0} &\al^{2i_1}& \ldots &\al^{2i_{\hs_{j+1}}}\\
\vdots &\vdots &\vdots &\ddots &\vdots \\
1&\al^{(\hs_{j+1}-1)i_0} &\al^{(\hs_{j+1}-1)i_1}& \ldots &\al^{(\hs_{j+1}-1)i_{\hs_{j+1}}}\\
\end{array}
\right).
$$

In particular,

\begin{eqnarray}
\label{ref}
\bigoplus_{s=0}^{\hs_{j+1}}\al^{ri_s}v_s&\,=\,&0\;\;{\rm for}\;\;0\leq
r\leq \hs_{j+1}-1.
\end{eqnarray}

Consider the $m\times n$ array of weight
$\left(\hs_{j+1}+1\right)\left(u_j+1\right)$ such that row $i_s$ equals
$v_s\,\uw$ for $0\leq s\leq \hs_{j+1}$, and the remaining rows are zero.
We will show that this array is in
$\C(n;k,\uu)$. Since each row of the array is in $\C_j$ by design, in
particular, it is in $\C_0$. Next,
according to~(\ref{eqGP1}) and~(\ref{eqGP2}), we have to show that

\begin{eqnarray}
\label{eqGP11b}
\bigoplus_{s=0}^{\hs_{j+1}}\al^{ri_s}\left(v_s\,\uw\right)&\in&
\C_{i}\;\;{\rm for}\;\; 1\leq i\leq t-1\;\; {\rm and}\;\; 0\leq r\leq \hs_{i}-1\\
\label{eqGP22}
\bigoplus_{s=0}^{\hs_{j+1}}\al^{ri_s}\left(v_s\,\uw\right)&=& 0\;\;{\rm for}\;\; 0\leq
r\leq m-k-1.
\end{eqnarray}

If $0\leq j\leq i-1$, then, $\hs_{j+1}\geq \hs_{i}$,
and, for $0\leq r\leq \hs_{i}-1$, by~(\ref{ref}),
\begin{eqnarray*}
\bigoplus_{s=0}^{\hs_{j+1}}\al^{ri_s}\left(v_s\,\uw\right)\;\;\,=\,\;\;
\left(\bigoplus_{s=0}^{\hs_{j+1}}\al^{ri_s}v_s\right)\,\uw&\,=\,
&0,\end{eqnarray*}
so~(\ref{eqGP11b}) follows. Since $m-k\leq\hs_{j+1}$, also
(\ref{eqGP22}) follows from~(\ref{ref}).

If $i\leq j\leq t-1$, then $\C_j\subseteq
\C_{i}$ and $\uw\in\C_{i}$,
so~(\ref{eqGP11b}) also follows in this case.

\qed

\begin{ex}
\label{ex9}
{\em
Consider the 3-level GPC code $\C(7;4,(1,1,3,4,4,4))$ of
Example~\ref{ex7}. According to Lemma~\ref{lemma1}, the locations
denoted by $E$ in the following arrays correspond to the non-zero
entries of arrays in $\C(7;4,(1,1,3,4,4,4))$:

$$
\begin{array}{lll}
\begin{array}{|c|c|c|c|c|c|c|}
\hline
\phantom{X}&E&\phantom{X}&E&\phantom{X}&\phantom{X}&\phantom{X}\\
\hline
\phantom{X}&E&\phantom{X}&E&\phantom{X}&\phantom{X}&\phantom{X}\\
\hline
\phantom{X}&\phantom{X}&\phantom{X}&\phantom{X}&\phantom{X}&\phantom{X}&\phantom{X}\\
\hline
\phantom{X}&E&\phantom{X}&E&\phantom{X}&\phantom{X}&\phantom{X}\\
\hline
\phantom{X}&E&\phantom{X}&E&\phantom{X}&\phantom{X}&\phantom{X}\\
\hline
\phantom{X}&E&\phantom{X}&E&\phantom{X}&\phantom{X}&\phantom{X}\\
\hline
\end{array}
&
\begin{array}{|c|c|c|c|c|c|c|}
\hline
\phantom{X}&E&E&\phantom{X}&E&\phantom{X}&E\\
\hline
\phantom{X}&\phantom{X}&\phantom{X}&\phantom{X}&\phantom{X}&\phantom{X}&\phantom{X}\\
\hline
\phantom{X}&E&E&\phantom{X}&E&\phantom{X}&E\\
\hline
\phantom{X}&E&E&\phantom{X}&E&\phantom{X}&E\\
\hline
\phantom{X}&\phantom{X}&\phantom{X}&\phantom{X}&\phantom{X}&\phantom{X}&\phantom{X}\\
\hline
\phantom{X}&E&E&\phantom{X}&E&\phantom{X}&E\\
\hline
\end{array}
&
\begin{array}{|c|c|c|c|c|c|c|}
\hline
\phantom{X}&\phantom{X}&\phantom{X}&\phantom{X}&\phantom{X}&\phantom{X}&\phantom{X}\\
\hline
E&\phantom{X}&E&\phantom{X}&E&E&E\\
\hline
\phantom{X}&\phantom{X}&\phantom{X}&\phantom{X}&\phantom{X}&\phantom{X}&\phantom{X}\\
\hline
\phantom{X}&\phantom{X}&\phantom{X}&\phantom{X}&\phantom{X}&\phantom{X}&\phantom{X}\\
\hline
E&\phantom{X}&E&\phantom{X}&E&E&E\\
\hline
E&\phantom{X}&E&\phantom{X}&E&E&E\\
\hline
\end{array}
\end{array}
$$

The arrays with erasures in locations $E$ above are uncorrectable,
since, provided the zero array was stored, the decoding cannot decide
between the zero array and the arrays with non-zero entries in the
locations $E$.

\qed
}
\end{ex}

Before stating the dimension $K$ of a $t$-level GPC code
$\C(n;k,\uu)$, we give an auxiliary general lemma.

\begin{lemma}
\label{lemma6}
{\em
Consider and $[n,k]$ code, and let $S\eq\{i_0,i_1,\ldots,i_{s-1}\}$,
where $0\,\leq \,i_0\,<\,i_1\,<\,\cdots \,<\,i_{s-1}\,\leq \,n-1$.
Assume that, given a codeword with erasures in $S$, the code can
correct such erasures, while, for any $i\not\in S$, erasures in
$S\cup \{i\}$ are not correctable. Then,
\begin{eqnarray}
\label{nminusk}
n-k&=&s.
\end{eqnarray}
}
\end{lemma}

\noindent\pf Since the erasures in $S$ are correctable, there are at
least $s$ linearly independent parity equations, so

\begin{eqnarray*}
n-k&\geq &s.
\end{eqnarray*}

Assume that $n-k\,>\,s$. Let $H$ be an $(n-k)\times n$ parity-check matrix of the code
such that the first $s$ rows of $H$ are used to correct the $s$
erasures in $S$, thus, the $s\times s$ submatrix consisting of those first $s$
rows and columns $i_0,i_1,\ldots,i_{s-1}$ is invertible.

Consider next the matrix consisting of the first $s+1$ rows in $H$.
By row operations, we can make the entries $i_0,i_1,\ldots,i_{s-1}$
in the $(s+1)$-th row equal to zero. Since the first $s+1$ rows of
$H$ have rank $s+1$, then there is a non-zero location $i$, $i\not\in
S$, in the $(s+1)$-th row. Thus, columns $S\cup\{i\}$ in the first
$s+1$ rows of $H$ are linearly
independent and hence erasures in $S\cup\{i\}$ are
correctable, a contradiction, so~(\ref{nminusk}) holds.

\qed

\begin{cor}
\label{cor00}
{\em
Consider the $t$-level GPC code $\C(n;k,\uu)$ as given by
Definition~\ref{defGPMDS}. Then, $\C(n;k,\uu)$ is an $[N,K]$ code, where
$N\eq mn$ and

\begin{eqnarray}
\label{eqK}
K&\eq &
kn-\left(\sum_{i=0}^{t-2}s_iu_i\right)-(s_{t-1}-m+k)u_{t-1}.
\end{eqnarray}

}
\end{cor}

\noindent\pf
Let $\hs_t\,=\,m-k$ and
$\hs_j$ be given by~(\ref{hsi}) for $0\leq j\leq t-1$.
Assume that the zero array is stored, and a received array $W$ has
erasures in the last $u_i$ entries of rows
$m-\hs_{i}$ to $m-\hs_{i+1}-1$ for $0\leq i\leq t-1$,
and in all the entries of rows $k$ 
to $m-1$. Thus, $W$ has a total
of $$n(m-k)+\left(\sum_{i=0}^{t-2}s_iu_i\right)+(s_{t-1}-m+k)u_{t-1}$$ erasures,
and by Theorem~\ref{theo2}, it will be correctly decoded as the zero codeword. 

Consider an array $V$ which coincides with $W$, except in one
location in which it has an extra erasure. 
We will show that any such $V$ is
uncorrectable, 
so, by Lemma~\ref{lemma6}, 
\begin{eqnarray*}
N-K&\,=\,&n(m-k)+\left(\sum_{i=0}^{t-2}s_iu_i\right)+(s_{t-1}+m+k)u_{t-1},
\end{eqnarray*}
which is equivalent to~(\ref{eqK}).

For each $(u',v')$ such that $(u',v')$ is not in the set of erasures
of $W$, define $i'$, $0\leq i'\leq t-1$, such that $m-\hs_{i'}\leq
u'\leq m-\hs_{i'+1}-1$, 
and let 
$Y^{(u',v')}\,=\,\left(y^{(u',v')}_{a,b}\right)_{0\leq a\leq m-1\atop
0\leq b\leq n-1}$  
be an array in $\C(n;k,\uu)$ whose non-zero coordinates
are in the intersection 
of rows $u',u'+1,\ldots,u'+\hs_{i'+1}$ 
and columns
$v',n-u_{i'},n-u_{i'}+1,\ldots,n-1$. Such a non-zero array exists
due to Lemma~\ref{lemma1}.

Assume that the extra erasure in $V$ is in location $(u,v)$, and if
$u_t\,=\,n$, define $j$, $0\leq j\leq t$, such that 
$n-u_{j}\leq v\leq n-u_{j-1}-1$.
Consider the arrays $Y^{(u',v)}$, where $u\leq u'\leq m-\hs_{j}-1$.
For each $u'$, $u<u'\leq m-\hs_{j}-1$, choose constants $c_{u'}$ such that
$$y^{(u,v)}_{u',v}\,\oplus\,\bigoplus_{z=u+1}^{u'}c_zy^{(z,v)}_{u',v}\,=\,0.$$
Then, defining
$$Y\,=\,\bigoplus_{z=u}^{m-\hs_j-1}Y^{(z,v)},$$ we can see that
$Y$ has a non-zero entry in $(u,v)$, while the remaining non-zero
entries are contained in the locations of the erasures of $W$. So,
array 
$V$ is uncorrectable, since it
can be decoded either as the zero array or as $Y$.
\qed

Theorem II.1 in~\cite{tk}, which corresponds to Corollary~2
in~\cite{bh}, is a special case of Corollary~\ref{cor00}.

The encoding is a special case of the decoding. For example, we may
place the parities at the end of the array in increasing order of
parities, as shown in Corollary~\ref{cor00}.
The parities are considered as erasures and may be obtained using the
triangulation
method described in Theorem~\ref{theo2}. The fact that the locations
of the erasures
are known allows for a simplification of the decoding algorithm. For
example, the triangulated matrix corresponding to the coefficients
of~(\ref{eqGP11t}) may be precomputed. We omit the
implementation details.

\begin{ex}
\label{ex10}
{\em
We illustrate the proof of Corollary~\ref{cor00} with the 3-level GPC
code\\ $\C(7;4,(1,1,3,4,4,4))$ of Examples~\ref{ex7} and~\ref{ex9}.
By Corollary~\ref{cor00}, this code is a $[42,19]$ code.
Following the proof of Corollary~\ref{cor00}, denote by $E$ the
erased locations in an array $W$:

\begin{eqnarray*}
W&=&
\begin{array}{|c|c|c|c|c|c|c|}
\hline
\phantom{X}&\phantom{X}&\phantom{X}&\phantom{X}&\phantom{X}&\phantom{X}&E\\
\hline
\phantom{X}&\phantom{X}&\phantom{X}&\phantom{X}&\phantom{X}&\phantom{X}&E\\
\hline
\phantom{X}&\phantom{X}&\phantom{X}&\phantom{X}&E&E&E\\
\hline
\phantom{X}&\phantom{X}&\phantom{X}&E&E&E&E\\
\hline
E&E&E&E&E&E&E\\
\hline
E&E&E&E&E&E&E\\
\hline
\end{array}
\end{eqnarray*}

If the non-erased locations of $W$ are zero, by Theorem~\ref{theo2}, the
array will be decoded as the zero array. Now, consider the array $V$
which has an
extra erasure in location $(u,v)\eq (0,1)$, rendering

\begin{eqnarray*}
V&=&
\begin{array}{|c|c|c|c|c|c|c|}
\hline
\phantom{X}&E&\phantom{X}&\phantom{X}&\phantom{X}&\phantom{X}&E\\
\hline
\phantom{X}&\phantom{X}&\phantom{X}&\phantom{X}&\phantom{X}&\phantom{X}&E\\
\hline
\phantom{X}&\phantom{X}&\phantom{X}&\phantom{X}&E&E&E\\
\hline
\phantom{X}&\phantom{X}&\phantom{X}&E&E&E&E\\
\hline
E&E&E&E&E&E&E\\
\hline
E&E&E&E&E&E&E\\
\hline
\end{array}
\end{eqnarray*}

Consider the following arrays $Y^{(u',1)}$, $0\leq u'\leq 3$, defined
as in Corollary~\ref{cor00}, whose non-zero entries are denoted
$y^{(u',1)}_{a,b}$ below:

\begin{eqnarray*}
Y^{(0,1)}&=&
\begin{array}{|c|c|c|c|c|c|c|}
\hline
\;0\;&y^{(0,1)}_{0,1}&\;0\;&\;0\;&\;0\;&\;0\;&y^{(0,1)}_{0,6}\\
\hline
\;0\;&y^{(0,1)}_{1,1}&\;0\;&\;0\;&\;0\;&\;0\;&y^{(0,1)}_{1,6}\\
\hline
\;0\;&y^{(0,1)}_{2,1}&\;0\;&\;0\;&\;0\;&\;0\;&y^{(0,1)}_{2,6}\\
\hline
\;0\;&y^{(0,1)}_{3,1}&\;0\;&\;0\;&\;0\;&\;0\;&y^{(0,1)}_{3,6}\\
\hline
\;0\;&y^{(0,1)}_{4,1}&\;0\;&\;0\;&\;0\;&\;0\;&y^{(0,1)}_{4,6}\\
\hline
0&0&0&0&0&0&0\\
\hline
\end{array}.
\end{eqnarray*}

\begin{eqnarray*}
Y^{(1,1)}&=&
\begin{array}{|c|c|c|c|c|c|c|}
\hline
0&0&0&0&0&0&0\\
\hline
\;0\;&y^{(1,1)}_{1,1}&\;0\;&\;0\;&\;0\;&\;0\;&y^{(1,1)}_{1,6}\\
\hline
\;0\;&y^{(1,1)}_{2,1}&\;0\;&\;0\;&\;0\;&\;0\;&y^{(1,1)}_{2,6}\\
\hline
\;0\;&y^{(1,1)}_{3,1}&\;0\;&\;0\;&\;0\;&\;0\;&y^{(1,1)}_{3,6}\\
\hline
\;0\;&y^{(1,1)}_{4,1}&\;0\;&\;0\;&\;0\;&\;0\;&y^{(1,1)}_{4,6}\\
\hline
\;0\;&y^{(1,1)}_{5,1}&\;0\;&\;0\;&\;0\;&\;0\;&y^{(1,1)}_{5,6}\\
\hline
\end{array}.
\end{eqnarray*}

\begin{eqnarray*}
Y^{(2,1)}&=&
\begin{array}{|c|c|c|c|c|c|c|}
\hline
0&0&0&0&0&0&0\\
\hline
0&0&0&0&0&0&0\\
\hline
\;0\;&y^{(2,1)}_{2,1}&\;0\;&\;0\;&y^{(2,1)}_{2,4}&y^{(2,1)}_{2,5}&y^{(2,1)}_{2,6}\\
\hline
\;0\;&y^{(2,1)}_{3,1}&\;0\;&\;0\;&y^{(2,1)}_{3,4}&y^{(2,1)}_{3,5}&y^{(2,1)}_{3,6}\\
\hline
\;0\;&y^{(2,1)}_{4,1}&\;0\;&\;0\;&y^{(2,1)}_{4,4}&y^{(2,1)}_{4,5}&y^{(2,1)}_{4,6}\\
\hline
\;0\;&y^{(2,1)}_{5,1}&\;0\;&\;0\;&y^{(2,1)}_{5,4}&y^{(2,1)}_{5,5}&y^{(2,1)}_{5,6}\\
\hline
\end{array}.
\end{eqnarray*}

\begin{eqnarray*}
Y^{(3,1)}&=&
\begin{array}{|c|c|c|c|c|c|c|}
\hline
0&0&0&0&0&0&0\\
\hline
0&0&0&0&0&0&0\\
\hline
0&0&0&0&0&0&0\\
\hline
\;0\;&y^{(3,1)}_{3,1}&\;0\;&y^{(3,1)}_{3,2}&y^{(3,1)}_{3,4}&y^{(3,1)}_{3,5}&y^{(3,1)}_{3,6}\\
\hline
\;0\;&y^{(3,1)}_{4,1}&\;0\;&y^{(3,1)}_{4,2}&y^{(3,1)}_{4,4}&y^{(3,1)}_{4,5}&y^{(3,1)}_{4,6}\\
\hline
\;0\;&y^{(3,1)}_{5,1}&\;0\;&y^{(3,1)}_{5,2}&y^{(3,1)}_{5,4}&y^{(3,1)}_{5,5}&y^{(3,1)}_{5,6}\\
\hline
\end{array}.
\end{eqnarray*}

Such arrays with non-zero entries exist by Lemma~\ref{lemma1} (see
also Example~\ref{ex9}).
We choose $c_1$, $c_2$ and $c_3$  
such that

\begin{eqnarray*}
y^{(0,1)}_{1,1}\xor c_1y^{(1,1)}_{1,1}&=&0\\
y^{(0,1)}_{2,1}\xor c_1y^{(1,1)}_{2,1}\xor c_2y^{(2,1)}_{2,1}&=&0\\
y^{(0,1)}_{3,1}\xor c_1y^{(1,1)}_{3,1}\xor c_2y^{(2,1)}_{3,1}\xor c_3y^{(3,1)}_{3,1}&=&0\\
\end{eqnarray*}

Then, defining $Y\,=\,Y^{(0,1)}\xor c_1Y^{(1,1)}\xor c_2Y^{(2,1)}\xor
c_3Y^{(3,1)}$, we see that

\begin{eqnarray*}
Y&=&
\begin{array}{|c|c|c|c|c|c|c|}
\hline
\;0\;&y^{(0,1)}_{0,1}&\;0\;&\;0\;&\;0\;&\;0\;&X\\
\hline
0&0&0&0&0&0&X\\
\hline
0&0&0&0&X&X&X\\
\hline
0&0&0&X&X&X&X\\
\hline
0&X&0&X&X&X&X\\
\hline
0&X&0&X&X&X&X\\
\hline
\end{array}
\end{eqnarray*}

Array $Y$ is non-zero since $y^{(0,1)}_{0,1}\neq 0$ (entries denoted
by $X$ may take any value). Array $V$
may be decoded either as the zero array or as $Y$, so it is
uncorrectable. Since we can make the same argument for any entry
$(u,v)$ not contained in the erasures of $W$, by Lemma~\ref{lemma6},
the number of parity
symbols is exactly 23 and the dimension of the code is 19.

\qed
}
\end{ex}


The following corollary extends Theorem~II.2 on $t$-level
II codes as stated in~\cite{tk} and proven as Corollary~3
in~\cite{bh}.
It also
generalizes the well known result that the minimum distance of a
product code is the product of the minimum distances of the two
component codes.

\begin{cor}
\label{cor11}
{\em
Consider the $t$-level GPC code $\C(n;k,\uu)$ as given
by~Definition~\ref{defGPMDS}. Then, if 
$\hs_t\eq m-k$ and $\hs_i$ is given by~(\ref{hsi})
for $0\leq i\leq t-1$, the minimum
distance of $\C(n;k,\uu)$ is

\begin{eqnarray}
\label{dist}
d&=&\min\left\{\left(\hs_{i+1}+1\right)\left(u_i+1\right)\;,\;0\leq
i\leq t-1\right\}.
\end{eqnarray}
}
\end{cor}

\noindent\pf
For each $i$ such that $0\leq i\leq t-1$, consider an array in $\C(n;k,\uu)$
that has $\hs_{i+1}$ rows with $u_i+1$ erasures each, one row
with $u_i$ erasures, and all the other entries are zero. By
Theorem~\ref{theo2}, such arrays will be corrected by
the code $\C(n;k,\uu)$ as the zero codeword, thus

\begin{eqnarray*}
d&\leq &\min\left\{\left(\hs_{i+1}+1\right)\left(u_i+1\right)\;,\;0\leq
i\leq t-1\right\}.
\end{eqnarray*}

On the other hand, by Lemma~\ref{lemma1}, for each $0\leq i\leq t-1$,
there is an array in $\C(n;k,\uu)$ of weight
$\left(\hs_{i+1}+1\right)\left(u_i+1\right)$, so

\begin{eqnarray*}
d&\geq &\min\left\{\left(\hs_{i+1}+1\right)\left(u_i+1\right)\;,\;0\leq
i\leq t-1\right\}
\end{eqnarray*}
and~(\ref{dist}) follows.

\qed

\begin{ex}
\label{ex8}
{\em
Consider the 3-level GPC code $\C(7;4,(1,1,3,4,4,4))$ of Example~\ref{ex7}.
According to Corollary~\ref{cor11}, since $m\eq 6$, $k\eq 4$, $u_0\eq
1$, $u_1\eq 3$, $u_2\eq 4$, $s_0\eq 2$, $s_1\eq 1$, $s_2\eq 3$ (and
hence, $\hs_3\eq m-k\eq 2$, $\hs_2\eq s_2\eq 3$, $\hs_1\eq s_1+s_2\eq 4$),
according to~(\ref{dist}), the minimum distance of this code is

\begin{eqnarray*}
d&=&
\min\left\{(5)(2)\,;(4)(4)\,;\,(3)(5)\right\}\;\eq\; 10.
\end{eqnarray*}

\qed
}
\end{ex}

Consider next a product code, such that the vertical code is an
$[m,k_0,m-k_0+1]$ code, and the horizontal code is an
$[n,k_1,n-k_1+1]$ code. In the notation of GPC codes, we denote this
1-level GPC code as
$\C(n;k_0,\overbrace{n-k_1,n-k_1,\ldots,n-k_1}^{m}\,)$ (see
Definition~\ref{defGPMDS} and Example~\ref{ex4}).
We can look at it also from the perspective of columns, and
then the code is a 1-level  GPC code 
$\C(m;k_1,\overbrace{m-k_0,m-k_0,\ldots,m-k_0}^{n}\,)$.
The following theorem generalizes this argument for a
$t$-level GPC code.

\begin{theo}
\label{theo3}
{\em
Consider an $m\times n$ array corresponding to a $\C(n;k,\uu)$
$t$-level GPC code as given by Definition~\ref{defGPMDS}.
Then, viewed as an $n\times m$ array on columns,
the code is a $\C(m;n-u_0,\uu')$ $t$-level GPC code, where
$\hs_t\eq m-k$, $\hs_i$ is given by~(\ref{hsi}) for $1\leq
i\leq t-1$, $u_t\eq n$,

\begin{eqnarray}
\label{equu'}
\uu' &=&
\left(\overbrace{u'_0,u'_0,\ldots,u'_0}^{s'_0},\overbrace{u'_1,u'_1,\ldots,u'_1}^{s'_1},\ldots,
\overbrace{u'_{t-1},u'_{t-1},\ldots,u'_{t-1}}^{s'_{t-1}}\right),
\end{eqnarray}
\begin{eqnarray}
\label{uprimeis}
u'_{t-i}\eq \hs_i\quad {\rm for}\quad 1\leq i\leq t,\quad s'_i\eq
u_{t-i}-u_{t-i-1}\quad {\rm for}\quad 0\leq i\leq
t-2\quad {\rm and}\quad s'_{t-1}\eq u_1.
\end{eqnarray}
}
\end{theo}

\noindent\pf Denote by $\uc^{\rm\bf  (H)}_i$, $0\leq i\leq m-1$, the rows of the
array, and by $\uc^{\rm\bf  (V)}_j$, $0\leq j\leq n-1$, the columns.
Specifically, if the array consists of symbols $(c_{i,j})_{0\leq
i\leq m-1\atop 0\leq j\leq n-1}$, then

\begin{eqnarray*}
\uc^{\rm\bf  (H)}_i&=&(c_{i,0},c_{i,1},\ldots,c_{i,n-1})\quad {\rm
for}\quad 0\leq i\leq m-1
\end{eqnarray*}
and
\begin{eqnarray*}
\uc^{\rm\bf  (V)}_j&=&(c_{0,j},c_{1,j},\ldots,c_{m-1,j})\quad {\rm
for}\quad 0\leq j\leq n-1.
\end{eqnarray*}

Consider the $t$ nested codes (on columns)
$\C'_{t-1}\subset\C'_{t-2}\subset\cdots\subset\C'_{0}$, where $\C'_i$ is an\\
$[m,m-u'_i,u'_i+1]$ code. A parity-check matrix of
$\C'_i$ is

\begin{eqnarray}
\label{Hprimei}
H'_i&=&\left(
\begin{array}{ccccc}
1&1&1&\ldots &1\\
1&\al &\al^2&\ldots &\al^{m-1}\\
1&\al^2 &\al^4&\ldots &\al^{2(m-1)}\\
\vdots &\vdots &\vdots &\ddots &\vdots \\
1&\al^{u'_{i}-1} &\al^{2(u'_{i}-1)}&\ldots &\al^{(m-1)(u'_{i}-1)}\\
\end{array}
\right)
\end{eqnarray}

In order to prove the theorem, according to
Definition~\ref{defGPMDS}, we have to prove that each
$\uc^{\rm\bf  (V)}_j\in\C'_{0}$, $0\leq j\leq n-1$, and

\begin{eqnarray}
\label{eqGP1'}
\bigoplus_{j=0}^{n-1}\al^{rj}\uc^{\rm\bf  (V)}_j&\in&
\C'_{i}\;\;{\rm for}\;\;
1\leq i\leq t-1\;\; {\rm and}\;\; 0\leq r\leq \hat{s'}_{i}-1\\
\label{eqGP2'}
\bigoplus_{j=0}^{n-1}\al^{ij}\uc^{\rm\bf (V)}_j&=& 0\;\;{\rm for}\;\; 0\leq
i\leq u_0-1.
\end{eqnarray}

$\C'_{0}$ is an $[m,m-u'_0,u'_0+1]$ code and by~(\ref{uprimeis}),
$u'_0\eq m-k$, so from~(\ref{eqGP2}), $\uc^{\rm\bf  (V)}_j\in\C'_{0}$.

Notice also that since each $\uc^{\rm\bf
(H)}_i\in\C_{0}$ for $0\leq i\leq m-1$, (\ref{eqGP2'}) follows.

Next we have to
prove~(\ref{eqGP1'}). In effect, (\ref{eqGP1'}) holds
if and only if, by~(\ref{Hprimei}),
\begin{eqnarray*}
\bigoplus_{v=0}^{m-1}\al^{uv}\bigoplus_{j=0}^{n-1}\al^{rj}c_{v,j}&=&
0\quad {\rm for}\quad 1\leq i\leq t-1,\;0\leq u\leq u'_{i}-1\;\; {\rm and}\;\; 0\leq r\leq \hat{s'}_{i}-1,
\end{eqnarray*}
if and only if, changing the summation order,
\begin{eqnarray*}
\bigoplus_{j=0}^{n-1}\al^{rj}\bigoplus_{v=0}^{m-1}\al^{uv}c_{v,j}&=&
0\quad {\rm for}\quad 1\leq i\leq t-1,\;0\leq u\leq u'_{i}-1\;\; {\rm and}\;\; 0\leq r\leq \hat{s'}_{i}-1,
\end{eqnarray*}
if and only if
\begin{eqnarray}
\label{hatC}
\bigoplus_{v=0}^{m-1}\al^{uv}\uc^{\rm\bf  (H)}_{v}&\in&\hat{\C}_i
\quad {\rm for}\quad 1\leq i\leq t-1,\;0\leq u\leq u'_{i}-1,
\end{eqnarray}
where a parity-check matrix for $\hat{\C}_i$ is given by

\begin{eqnarray}
\label{hatH}
\hat{H}_i&=&\left(
\begin{array}{ccccc}
1&1&1&\ldots &1\\
1&\al &\al^2&\ldots &\al^{n-1}\\
1&\al^2 &\al^4&\ldots &\al^{2(n-1)}\\
\vdots &\vdots &\vdots &\ddots &\vdots \\
1&\al^{\hat{s'}_{i}-1} &\al^{2(\hat{s'}_{i}-1)}&\ldots &\al^{(n-1)(\hat{s'}_{i}-1)}\\
\end{array}
\right).
\end{eqnarray}

By~(\ref{hsi}) and~(\ref{uprimeis}) , 
$$\hat{s'}_{i}\eq \sum_{z=i}^{t-1}s'_z\eq
\left(\sum_{z=i}^{t-2}u_{t-z}-u_{t-z-1}\right)+u_1\eq u_{t-i},$$
so, by~(\ref{hatH}) and~(\ref{Hi}), $\hat{H}_i\eq H_{t-i}$ and hence
$\hat{\C}_i\eq \C_{t-i}$. By~(\ref{uprimeis}), $u'_{i}\eq \hs_{t-i}$,
so~(\ref{hatC}) becomes

\begin{eqnarray*}
\bigoplus_{v=0}^{m-1}\al^{uv}\uc^{\rm\bf  (H)}_{v}&\in& \C_{t-i}\;\;{\rm for}\;\; 1\leq
i\leq t-1\;\; {\rm and}\;\; 0\leq u\leq \hs_{t-i}-1,\\
\end{eqnarray*}
which is equivalent to~(\ref{eqGP1}) and thus~(\ref{eqGP1'})
and~(\ref{eqGP1}) are equivalent, 
completing the proof.

\qed

\begin{ex}
\label{ex11}
{\em
Consider the 3-level GPC code $\C(7;5,(1,1,3,3,5,5))$.
According to Theorem~\ref{theo3},
this code is also a 3-level GPC code $\C(6;6,(1,1,2,2,4,4,4))$
consisting of $7\times 6$ arrays, so, according to
Theorem~\ref{theo2}, it can correct any column
with one erasure, up to two columns with 2 erasures, up to 2
columns with 4 erasures and up to one erased column. This allows for correction
of erasures that cannot be handled by the correction on rows. For
example, consider the following array, where the erasures are denoted
by $E$:

$$
\begin{array}{|c|c|c|c|c|c|c|}
\hline
E&\phantom{X}&\phantom{X}&\phantom{X}&E&\phantom{X}&\phantom{X}\\
\hline
E&E&\phantom{X}&\phantom{X}&\phantom{X}&\phantom{X}&\phantom{X}\\
\hline
\phantom{X}&E&E&\phantom{X}&\phantom{X}&\phantom{X}&\phantom{X}\\
\hline
\phantom{X}&\phantom{X}&E&E&\phantom{X}&\phantom{X}&\phantom{X}\\
\hline
\phantom{X}&\phantom{X}&\phantom{X}&E&E&\phantom{X}&\phantom{X}\\
\hline
\phantom{X}&\phantom{X}&\phantom{X}&\phantom{X}&\phantom{X}&\phantom{X}&\phantom{X}\\
\hline
\end{array}
$$

The erasures cannot be decoded by the horizontal code
$\C(7;5,(1,1,3,3,5,5))$, but they can certainly be handled by the
vertical code $\C(6;6,(1,1,2,2,4,4,4))$.

\qed
}
\end{ex}

Example~\ref{ex11} suggests an expansion of the decoding algorithm as
given in the proof by triangulation of Theorem~\ref{theo2}: given a
$t$-level GPC code $\C(n;k,\uu)$, each time there are erasures we
apply the decoding algorithm on rows as described in
Theorem~\ref{theo2}. If after this process there are still erasures
remaining, we apply the decoding algorithm on columns for the
$\C(m;m-u_0,\uu')$ $t$-level GPC code as determined by
Theorem~\ref{theo3}. The method extends the decoding method of
product codes, in which erasures are iteratively corrected by both
codes, until they are either corrected or an uncorrectable
pattern remains.

Let us point out that the decoding algorithm can be adapted to handle
errors together with erasures, but we omit its description here.

\section{Extended Product Codes and Optimality Issues}
\label{optimality}
The $t$-level GPC codes $\C(n;k,\uu)$ described in Section~\ref{GP}
are a special case of product codes with some extra (global)
parities. Let us call an extended product (EPC) code such a code, and
denote it by $EP(m,v;n,h;g)$, where $v$ is the number of vertical
parities, $h$ the number of horizontal parities, and $g$ the number
of global parities. 
For example, the 3-level GPC code $\C(7;4,(1,1,3,4,4,4))$ of
Examples~\ref{ex7}, \ref{ex9} and~\ref{ex10} is an $EP(6,2;7,1;5)$,
while the 3-level GPC code $\C(7;5,(1,1,3,3,5,5))$ of
Example~\ref{ex11} is an $EP(6,1;7,1;8)$.

The next lemma gives an upper bound on the minimum distance of an
$EP(m,v;n,h;g)$ code.

\begin{lemma}
\label{lemma2}
{\em
Let $d(m,v;n,h;g)$ be the minimum distance of an $EP(m,v;n,h;g)$ code. Then,
\begin{eqnarray}
\label{dvhg}
d(m,v;n,h;g)\,\leq \,\min\{d(v,h,g\,;\,a)\,:\,
\lceil (g+1)/(m-v)\rceil\leq a\leq \min\{g+1\,,\,n-h\}\}, 
\end{eqnarray}
where, 
if $b\,=\,\lfloor (g+1)/a\rfloor$ and $r\,=\,g+1-ab$, then 

\begin{eqnarray}
\label{dvhga0}
d(v,h,g;a)&\,=\,&(v+b)(h+a)\quad {\rm for}\quad r\,=\,0
\end{eqnarray}
and

\begin{eqnarray}
\label{dvhgar}
d(v,h,g;a)&\,=\,&(v+b)(h+a)+h+r\quad {\rm for}\quad r\,\neq\,0.
\end{eqnarray}
}
\end{lemma}

\noindent\pf Assume first that $r\,=\,0$, the zero array is stored, and the received
array has the locations $(i,j)$ erased, where, by~(\ref{dvhg}),
$$0\leq i\leq v+b-1\leq v+{g+1\over (g+1)/(m-v)}-1\,=\ m-1$$ and 
$$0\leq j\leq h+a-1\leq h+(n-h)-1\,=\,n-1.$$ In particular, since $ab\,=\,g+1$, there are
$(v+b)(h+a)\,=\,vh+va+hb+g+1$ erasures. We
argue that such a received array is uncorrectable, which would
prove~(\ref{dvhg}) when $d(v,h,g;a)$ satisfies~(\ref{dvhga0}). Notice that we
have $h(v+b)$ horizontal parities and $v(h+a)$ vertical parities
corresponding to the
product code in order to correct the $(v+b)(h+a)$ erasures, but $hv$ of such
parities are dependent, so that leaves us with a total of
$h(v+b)+v(h+a)-hv\,=\,hv+hb+va$ parities corresponding to the
product code. In addition, $g$ global parities can be used,
giving a total of $hv+hb+va+g$ parities,
insufficient to correct the $hv+hb+va+g+1$ erasures.

Similarly, assume that $r\neq 0$, the zero array is stored, and the received
array has the locations $(i,j)$ erased, where
$0\leq i\leq v+b-1$, $0\leq j\leq h+a-1\leq n-1$, and in addition,
locations $(v+b,j)$ are also erased, where $0\leq j\leq h+r-1$.
Observe that all the erasures are within the array. In effect, 
since $a$ does not divide $g+1$,


$$v+b\,=\,v+\left\lfloor{g+1\over a}\right\rfloor\,<\,v+{g+1\over a}\,\leq\,
v+{g+1\over \lceil(g+1)/(m-v)\rceil}\,\leq\,v+(m-v)\,=\,m.$$ 
In particular, since $ab\,=\,g+1-r$, there are
$(v+b)(h+a)+h+r\,=\,hv+va+hb+g+h+1$ erasures. We will show that such a
received array is uncorrectable, which would 
prove~(\ref{dvhg}) when $d(v,h,g;a)$ satisfies~(\ref{dvhgar}). Notice that we
have $h(v+b+1)$ horizontal parities and $v(h+a)$ vertical parities
corresponding to the
product code in order to correct such patterns, but since, as before,
$hv$ of such parities are dependent, that gives a total of
$h(v+b+1)+v(h+a)-hv\,=\,hv+hb+va+h$ parities corresponding to the
product code. In addition, $g$ global parities can be used,
giving a total of $hv+hb+va+h+g$ parities,
insufficient to correct $hv+va+hb+g+h+1$ erasures.

\qed

\begin{ex}
\label{ex12}
{\em Consider an $EP(7,2;8,3;3)$ code and
let $d(7,2;8,3;3)$ be its minimum distance.
According to~(\ref{dvhg}),

\begin{eqnarray*}
d(7,2;8,3;3)&\leq &\min\{d(2,3,3;a)\,:\,1\leq a\leq 4\},
\end{eqnarray*}
where, according to~(\ref{dvhga0}) and~(\ref{dvhgar}),

\begin{eqnarray*}
d(2,3,3;1)&\eq &24\\
d(2,3,3;2)&\eq &20\\
d(2,3,3;3)&\eq &22\\
d(2,3,3;4)&\eq &21,\\
\end{eqnarray*}
so
\begin{eqnarray*}
d(7,2;8,3;3)&\leq &20.
\end{eqnarray*}

Following the proof of Lemma~\ref{lemma2}, a pattern of 20
uncorrectable erasures is given by

$$
\begin{array}{|c|c|c|c|c|c|c|c|}
\hline
E&E&E&E&E&\phantom{X}&\phantom{X}&\phantom{X}\\
\hline
E&E&E&E&E&\phantom{X}&\phantom{X}&\phantom{X}\\
\hline
E&E&E&E&E&\phantom{X}&\phantom{X}&\phantom{X}\\
\hline
E&E&E&E&E&\phantom{X}&\phantom{X}&\phantom{X}\\
\hline
\phantom{X}&\phantom{X}&\phantom{X}&\phantom{X}&\phantom{X}&\phantom{X}&\phantom{X}&\phantom{X}\\
\hline
\phantom{X}&\phantom{X}&\phantom{X}&\phantom{X}&\phantom{X}&\phantom{X}&\phantom{X}&\phantom{X}\\
\hline
\phantom{X}&\phantom{X}&\phantom{X}&\phantom{X}&\phantom{X}&\phantom{X}&\phantom{X}&\phantom{X}\\
\hline
\end{array}
$$

\qed
}
\end{ex}

We will say that an $EP(m,v;n,h;g)$ code is optimal if it meets
bound~(\ref{dvhg}) with equality. We will devote the rest of this
section to presenting some special cases of optimal $EP(m,v;n,h;g)$ codes.
We believe that there are optimal $EP(m,v;n,h;g)$ codes for any
choice of parameters, but the subject requires further research.

\begin{lemma}
\label{lemma4}
{\rm
Consider the 2-level GPC code $\C(n;k_0,\uu)$ as given by Definition~\ref{defGPMDS}, where
$$\uu\eq
(\overbrace{k_1,k_1,\ldots,k_1}^{m-k_0-1},\overbrace{k_1+1,k_1+1,\ldots,k_1+1}^{k_0+1}).$$
Then, $\C(n;k_0,\uu)$ is an optimal $EP(m,m-k_0;n,n-k_1;1)$ code.
}
\end{lemma}

\noindent\pf It is clear that $\C(n;k_0,\uu)$ is
an $EP(m,m-k_0;n,n-k_1;1)$ code.
By Corollary~\ref{cor11}, the minimum distance of this code is
\begin{eqnarray}
\label{d1}
d&\eq &\min\{(m-k_0+1)(n-k_1+2)\;,\;(n-k_1+1)(m-k_0+2)\}.
\end{eqnarray}

But the right hand side of~(\ref{d1}) coincides with the right hand
side of bound~(\ref{dvhg}), showing that when
$g\eq 1$, the bound is tight.

\qed

Notice that, in particular, if $m-k_0\eq n-k_1\eq 1$ (single
parity horizontal and vertical 
codes), then~(\ref{d1}) gives $d\eq 6$, as claimed in Example~\ref{ex5}.

Let us examine now the case of $EP(m,1;n,1;2)$ codes, where $m,n\geq 3$. In this case,
bound~(\ref{dvhg}) gives

\begin{eqnarray}
\label{d2}
d(m,1;n,1;2)&\leq &8.
\end{eqnarray}

Consider for example a 2-level GPC code
$\C(n;m-1,(\overbrace{1,1,\ldots,1}^{m-2}),3,3)$ or a 2-level GPC code
$\C(n;m-1,(\overbrace{1,1,\ldots,1}^{m-3}),2,2,2)$.
These are the only cases of GPC codes that are $EP(m,1;n,1;2)$ codes.
In both cases, according to Corollary~\ref{cor11}, the minimum
distance is 6, so bound~(\ref{d2}) is not met.

However, bound~(\ref{d2}) is tight, and to show this we present an
$EP(m,1;n,1;2)$ code with minimum distance 8.
The construction is related to the
PMDS constructions in~\cite{bpsy}, and we pay the price of
requiring a larger finite field than for GPC codes.

Let $GF(2^b)$ be a finite field and $\al$ and element in $GF(2^b)$
such that $mn\,\leq\,\cO(\al)$, where $\cO(\al)$ denotes the order of $\al$.
Consider the 
parity-check matrix $\cH_2$ 
given by

\begin{eqnarray}
\label{pc0}
\cH_2&=&\left(
\begin{array}{c}
I_m\otimes (\overbrace{1,1,\ldots,1}^n)\\
(\overbrace{1,1,\ldots,1}^m)\otimes I_n\\
\hline
\begin{array}{ccccc}
1&\al&\al^2&\ldots &\al^{mn-1}\\
1&\al^{-1}&\al^{-2}&\ldots &\al^{-(mn-1)}\\
\end{array}
\end{array}
\right),
\end{eqnarray}
where $I_m$ denotes the $m\times m$ identity matrix and $\otimes$ the
Kronecker product~\cite{ms} of two matrices. Notice that

$$
\left(
\begin{array}{c}
I_m\otimes (\overbrace{1,1,\ldots,1}^n)\\
(\overbrace{1,1,\ldots,1}^m)\otimes I_n\\
\end{array}
\right)
$$
corresponds to the parity-check matrix of the product code with
single parity in rows and columns. We denote the matrix
in~(\ref{pc0}) as $\cH_2$ to indicate that two extra parities are added
to the product code.

The following lemma gives the minimum distance of the code whose
parity-check matrix is $\cH_2$.

\begin{lemma}
\label{lemma3}
{\em
Consider the $EP(m,1;n,1;2)$ code whose parity-check matrix $\cH_2$ is
given by~(\ref{pc0}), $m,n\geq 3$ and $mn\,\leq\,\cO(\al)$. Then, the
code has minimum distance 8. 
}
\end{lemma}

\noindent\pf We have to prove that any 7 erasures can be corrected.

First assume that
there are six erasures in locations $(i_0,j_0)$, $(i_0,j_1)$, $(i_0,j_2)$, $(i_1,j_0)$,
$(i_1,j_1)$ and  $(i_1,j_2)$, where $0\leq i_0<i_1\leq m-1$ and $0\leq j_0<j_1<j_2\leq n-1$
or $(i_0,j_0)$, $(i_0,j_1)$, $(i_1,j_0)$, $(i_1,j_1)$,
$(i_2,j_0)$ and  $(i_2,j_1)$, where $0\leq i_0<i_1<i_2\leq m-1$ and $0\leq j_0<j_1\leq n-1$,
and a seventh erasure in any other location. This seventh erasure
can be corrected using either horizontal or vertical parities, thus,
it is enough to prove that the two situations of six erasures
described above are correctable.
For example, using $5\times 5$ arrays, these two
situations are illustrated below:

\vspace{.4cm}

\centerline{
\begin{tabular}{cc}
\begin{tabular}{|c|c|c|c|c|}
\hline
E&E&&E&\\
\hline
&&\phantom{0}&&\\
\hline
&&\phantom{0}&&\\
\hline
E&E&&E&\phantom{0}\\
\hline
&&\phantom{0}&&\\
\hline
\end{tabular}
&
\begin{tabular}{|c|c|c|c|c|}
\hline
\phantom{0}&E&&E&\\
\hline
&&\phantom{0}&&\\
\hline
&&\phantom{0}&&\\
\hline
&E&&E&\phantom{0}\\
\hline
&E&\phantom{0}&E&\\
\hline
\end{tabular}
\end{tabular}
}

\vspace{.4cm}

Consider the first case, as illustrated by the array in the left.
It suffices to prove, using the
parity-check matrix as given by~(\ref{pc0}) , that the
$6\times 6$ matrix

$$
\left(
\begin{array}{cccccc}
1&1&1&0&0&0\\
0&0&0&1&1&1\\
0&1&0&0&1&0\\
0&0&1&0&0&1\\
\al^{i_0n+j_0}&\al^{i_0n+j_1}&\al^{i_0n+j_2}&\al^{i_1n+j_0}&\al^{i_1n+j_1}&\al^{i_1n+j_2}\\
\al^{-i_0n-j_0}&\al^{-i_0n-j_1}&\al^{-i_0n-j_2}&\al^{-i_1n-j_0}&\al^{-i_1n-j_1}&\al^{-i_1n-j_2}\\
\end{array}
\right)
$$
is invertible. Redefining $i\la i_1-i_0$, $j_1\la j_1-j_0$ and $j_2\la j_2-j_0$, where
now $1\leq i\leq m-1$ and $1\leq j_1<j_2\leq n-1$, this
matrix is invertible if and only if matrix

$$
\left(
\begin{array}{cccccc}
1&1&1&0&0&0\\
0&0&0&1&1&1\\
0&1&0&0&1&0\\
0&0&1&0&0&1\\
1&\al^{j_1}&\al^{j_2}&\al^{in}&\al^{in+j_1}&\al^{in+j_2}\\
1&\al^{-j_1}&\al^{-j_2}&\al^{-in}&\al^{-in-j_1}&\al^{-in-j_2}\\
\end{array}
\right)
$$
is invertible. This matrix is invertible if and only if the $5\times
5$ matrix

$$
\left(
\begin{array}{ccccc}
1&0&0&1&0\\
1\xor\al^{j_1}&1\xor\al^{j_2}&\al^{in}&\al^{in+j_1}&\al^{in+j_2}\\
1\xor\al^{-j_1}&1\xor\al^{-j_2}&\al^{-in}&\al^{-in-j_1}&\al^{-in-j_2}\\
0&1&0&0&1\\
0&0&1&1&1\\
\end{array}
\right)
$$
is invertible, if and only if the $4\times 4$ matrix

$$
\left(
\begin{array}{cccc}
1&0&0&1\\
1\xor\al^{j_2}&\al^{in}&1\xor\al^{j_1}\xor\al^{in+j_1}&\al^{in+j_2}\\
1\xor\al^{-j_2}&\al^{-in}&1\xor\al^{-j_1}\xor\al^{-in-j_1}&\al^{-in-j_2}\\
0&1&1&1\\
\end{array}
\right)
$$
is invertible, if and only if the $3\times 3$ matrix

$$
\left(
\begin{array}{ccc}
1&1&1\\
\al^{in}&1\xor\al^{j_1}\xor\al^{in+j_1}&1\xor\al^{j_2}\xor\al^{in+j_2}\\
\al^{-in}&1\xor\al^{-j_1}\xor\al^{-in-j_1}&1\xor\al^{-j_2}\xor\al^{-in-j_2}\\
\end{array}
\right)
$$
is invertible, if and only if the $2\times 2$ matrix
\begin{eqnarray*}
\left(
\begin{array}{ccc}
(1\xor\al^{j_1})(1\xor\al^{in})&(1\xor\al^{j_2})(1\xor\al^{in})\\
(1\xor\al^{-j_1})(1\xor\al^{-in})&(1\xor\al^{-j_2})(1\xor\al^{-in})\\
\end{array}
\right)
\end{eqnarray*}
is invertible, if and only if, since $1\xor\al^{in}\neq 0$,
\begin{eqnarray*}
\left(
\begin{array}{ccc}
1\xor\al^{j_1}&1\xor\al^{j_2}\\
1\xor\al^{-j_1}&1\xor\al^{-j_2}\\
\end{array}
\right)&=&
\left(
\begin{array}{ccc}
1\xor\al^{j_1}&1\xor\al^{j_2}\\
\al^{-j_1}(1\xor\al^{j_1})&\al^{-j_2}(1\xor\al^{j_2})\\
\end{array}
\right)
\end{eqnarray*}
is invertible, if and only if, since $1\xor\al^{j_1}\neq 0$ and
$1\xor\al^{j_2}\neq 0$,
$$
\left(
\begin{array}{ccc}
1&1\\
\al^{-j_1}&\al^{-j_2}\\
\end{array}
\right)
$$
is invertible, if and only if $\al^{j_1}\neq \al^{j_2}$, which is the
case since $1\leq j_1<j_2\leq n-1<\cO(\al)$.

Consider now the second case, then, again by~(\ref{pc0}),
we have to prove that the $6\times 6$ matrix

$$
\left(
\begin{array}{cccccc}
1&1&0&0&0&0\\
0&0&1&1&0&0\\
0&0&0&0&1&1\\
0&1&0&1&0&1\\
\al^{i_0n+j_0}&\al^{i_0n+j_1}&\al^{i_1n+j_0}&\al^{i_1n+j_1}&\al^{i_2n+j_0}&\al^{i_2n+j_1}\\
\al^{-i_0n-j_0}&\al^{-i_0n-j_1}&\al^{-i_1n-j_0}&\al^{-i_1n-j_1}&\al^{-i_2n-j_0}&\al^{-i_2n-j_1}\\
\end{array}
\right)
$$
is invertible.

Redefining $i_1\la i_1-i_0$, $i_2\la i_2-i_0$ and $j\la j_1-j_0$,
where $1\leq i_1<i_2\leq m-1$ and $1\leq j\leq n-1$, the
matrix above is invertible if and only if the matrix

$$
\left(
\begin{array}{cccccc}
1&1&0&0&0&0\\
1&\al^{j}&\al^{i_1n}&\al^{i_1n+j}&\al^{i_2n}&\al^{i_2n+j}\\
1&\al^{-j}&\al^{-i_1n}&\al^{-i_1n-j}&\al^{-i_2n}&\al^{-i_2n-j}\\
0&0&1&1&0&0\\
0&0&0&0&1&1\\
0&1&0&1&0&1\\
\end{array}
\right)
$$
is invertible. Proceeding with Gaussian elimination like in the
previous case, this matrix is invertible if and only if the $2\time
2$ matrix

$$
\left(
\begin{array}{ccccc}
(1\xor \al^{j})(1\xor \al^{i_1n})&(1\xor\al^{j})(1\xor \al^{i_2n})\\
\al^{-j-i_1n}(1\xor \al^{j})(1\xor \al^{i_1n})&\al^{-j-i_2n}(1\xor\al^{j})(1\xor \al^{i_2n})\\
\end{array}
\right)
$$
is invertible, if and only if $\al^{i_1n}\neq \al^{i_2n}$, which is
the case since $$1\leq i_1n<i_2n\leq (m-1)n<\cO(\al).$$

Next, assume that there are seven erasures, such that each row and
column has at least two erasures. This can only happen if one row
(column) has three erasures and two rows (columns) have two erasures.
The situation is illustrated by the two cases below:

\vspace{.4cm}

\centerline{
\begin{tabular}{cc}
\begin{tabular}{|c|c|c|c|c|}
\hline
E&E &&&\\
\hline
&&\phantom{0}&&\\
\hline
&&\phantom{0}&&\\
\hline
E&E &&E&\phantom{0}\\
\hline
&E&\phantom{0}&E&\\
\hline
\end{tabular}
&
\begin{tabular}{|c|c|c|c|c|}
\hline
\phantom{0}&E&E& &\\
\hline
&&\phantom{0}&&\\
\hline
&&\phantom{0}&&\\
\hline
&E&E&E&\phantom{0}\\
\hline
&E&\phantom{0}&E&\\
\hline
\end{tabular}
\end{tabular}
}

\vspace{.4cm}

Let
$i_0$ be the row with three erasures, and $j_0$ the column with
three erasures, while $j_1<j_2$ and $i_1$ is such that erasures are
in $(i_1,j_0)$ and $(i_1,j_1)$ so the remaining two erasures are in
$(i_2,j_0)$ and $(i_2,j_2)$.
It suffices to prove, using the
parity-check matrix as given by~(\ref{pc0}), that the
$7\times 7$ matrix

$$
\left(
\begin{array}{ccccccc}
1&1&1&0&0&0&0\\
0&0&0&1&1&0&0\\
0&0&0&0&0&1&1\\
0&1&0&0&1&0&0\\
0&0&1&0&0&0&1\\
\al^{i_0n+j_0}&\al^{i_0n+j_1}&\al^{i_0n+j_2}&\al^{i_1n+j_0}&\al^{i_1n+j_1}&\al^{i_2n+j_0}&\al^{i_2n+j_2}\\
\al^{-i_0n-j_0}&\al^{-i_0n-j_1}&\al^{-i_0n-j_2}&\al^{-i_1n-j_0}&\al^{-i_1n-j_1}&\al^{-i_2n-j_0}&\al^{-i_2n-j_2}\\
\end{array}
\right)
$$
is invertible.

Redefining $i_1\la i_1-i_0$, $i_2\la i_2-i_0$, $j_1\la j_1-j_0$ and $j_2\la j_2-j_0$,
the matrix above is invertible if and only if the matrix

$$
\left(
\begin{array}{ccccccc}
1&1&1&0&0&0&0\\
0&0&0&1&1&0&0\\
0&0&0&0&0&1&1\\
0&1&0&0&1&0&0\\
0&0&1&0&0&0&1\\
1&\al^{j_1}&\al^{j_2}&\al^{i_1n}&\al^{i_1n+j_1}&\al^{i_2n}&\al^{i_2n+j_2}\\
1&\al^{-j_1}&\al^{-j_2}&\al^{-i_1n}&\al^{-i_1n-j_1}&\al^{-i_2n}&\al^{-i_2n-j_2}\\
\end{array}
\right)
$$
is invertible, if and only if, doing Gaussian elimination like in the
other two cases, the $2\times 2$ matrix

$$
\left(
\begin{array}{cc}
(1\xor\al^{j_1})(1\xor\al^{i_1n})&(1\xor\al^{j_2})(1\xor\al^{i_2n})\\
(1\xor\al^{-j_1})(1\xor\al^{-i_1n})&(1\xor\al^{-j_2})(1\xor\al^{-i_2n})\\
\end{array}
\right)
$$
is invertible, if and only if, since $1\xor\al^{j_1}$,
$1\xor\al^{i_1n}$, $1\xor\al^{j_2}$ and $1\xor\al^{i_2n}$ are non-zero,

$$
\left(
\begin{array}{cc}
1&1\\
\al^{-i_1n-j_1}&\al^{-i_2n-j_2}\\
\end{array}
\right)
$$
is invertible, if and only if, computing the determinant,

\begin{eqnarray*}
\al^{i_1n+j_1}&\neq &\al^{i_2n+j_2}
\end{eqnarray*}
which is the case
since $mn\,\leq\,\cO(\al)$, thus

\begin{eqnarray*}
(i_2-i_1)n+j_2-j_1 &\not\equiv &0\;(\bmod\;\cO(\al)).
\end{eqnarray*}

\qed

Lemma~\ref{lemma3} shows that the code given by parity-check matrix
$\cH_2$ meets bound~(\ref{d2}) with equality, something that could not be
achieved by GPC codes with two global parities. 

Consider the 3-level GPC code
$\C(n;m-1,\uu)$, where
$$\uu\eq (\overbrace{1,1,\ldots,1}^{m-3}),2,3,3).$$
This is an
$EP(n,1;m,1;3)$ code. According to Corollary~\ref{cor11},
$\C(n;m-1,\uu)$ has minimum
distance 8, the same as the code given by parity-check matrix $\cH_2$,
at the price of an extra parity. However, there is a tradeoff: the
size of the field required by $\C(n;m-1,\uu)$ is
greater than $\max\{m;n\}$, while the field required by the code
whose parity-check matrix is $\cH_2$ must have size greater than
$mn$. Also, by Theorem~\ref{theo2}, $\C(n;m-1,\uu)$ can
correct 8 erasures involving two rows with 3 erasures and one row
with two erasures, like for example
\begin{center}
\begin{tabular}{|c|c|c|c|c|}
\hline
E&E&&E&\\
\hline
&&\phantom{0}&&\\
\hline
E&E&\phantom{0}&&\\
\hline
E&E&&E&\phantom{0}\\
\hline
&&\phantom{0}&&\\
\hline
\end{tabular}
\end{center}
The code generated by $\cH_2$ is unable to correct such pattern since
it does not have enough parities, so
even if both codes have the same minimum distance,
$\C(n;m-1,\uu)$ can correct more
erasure patterns. These tradeoffs need to be evaluated when
implementation is considered.

Let us finish this section with the case of $EP(n,1;m,1;3)$ codes.
Bound~(\ref{dvhg}) gives

\begin{eqnarray}
\label{d3s}
d(m,1;n,1;3)&\leq &9.
\end{eqnarray}

The next question is if bound~(\ref{d3s}) is tight. The answer is
yes.

As in the case of two global parities, let $GF(2^b)$ be a finite
field and let $\al$ be an element in $GF(2^b)$ such that $mn\,\leq\,\cO (\al)$.
Consider the parity-check matrix $\cH_3$ given by

\begin{eqnarray}
\label{pc3}
\cH_3&=&\left(
\begin{array}{c}
I_m\otimes (\overbrace{1,1,\ldots,1}^n)\\
(\overbrace{1,1,\ldots,1}^m)\otimes I_n\\
\hline
\begin{array}{ccccc}
1&\al&\al^2&\ldots &\al^{mn-1}\\
1&\al^{-1}&\al^{-2}&\ldots &\al^{-(mn-1)}\\
1&\al^{2}&\al^{4}&\ldots &\al^{2(mn-1)}\\
\end{array}
\end{array}
\right).
\end{eqnarray}

Notice that $\cH_2$ as given by~(\ref{pc0}) consists of the first
$m+n+2$ rows of $\cH_3$.
The following lemma gives the minimum distance of these codes under a
certain condition.

\begin{lemma}
\label{lemma5}
{\em
The $EP(m,1;n,1;3)$ code, where $m,n\geq 3$ and $mn\,\leq\, \cO(\al)$
whose parity-check matrix $\cH_3$ is given by~(\ref{pc3}), has minimum
distance 9 if and only if, for any $i_1,i_2\neq 0$, 
$1\leq i_1\leq m-1$,
$1\leq |i_2|\leq m-1$
and $j_1,j_2\neq 0$,
$1\leq j_1\leq n-1$,
$1\leq |j_2|\leq n-1$,

\begin{eqnarray}
\label{inv1}
1\xor\al^{-j_1}\xor\al^{-i_2n+j_2}\xor\al^{-(i_2-i_1)n+j_2}\neq 0.
\end{eqnarray}
}
\end{lemma}

\noindent\pf We have to prove that any 8 erasures are going to be
corrected under condition~(\ref{inv1}).

Assume that there are 8 erasures, such that each row and
column has at least two erasures.
There are three situations under which this can happen:

\begin{enumerate}

\item Two rows have four erasures and four columns have two erasures.

\item Four rows have two erasures and two columns have four erasures.

\item Two rows (columns) have three erasures and one row (column) has two erasures.

\end{enumerate}

The situation is illustrated by the four cases below. The first array
illustrates case 1, the second array illustrates case 2, and the
third and fourth arrays illustrate case 3.

\vspace{.4cm}

\centerline{
\begin{tabular}{cccc}
\begin{tabular}{|c|c|c|c|c|}
\hline
E&E &\phantom{E}&E&E\\
\hline
&&\phantom{E}&&\\
\hline
&&\phantom{E}&&\\
\hline
E&E &&E&E\\
\hline
&&\phantom{E}&&\\
\hline
\end{tabular}
&
\begin{tabular}{|c|c|c|c|c|}
\hline
\phantom{E}&E&E&\phantom{E} &\\
\hline
&E&E&&\\
\hline
&&\phantom{E}&&\\
\hline
&E&E&&\phantom{E}\\
\hline
&E&E&&\\
\hline
\end{tabular}
&
\begin{tabular}{|c|c|c|c|c|}
\hline
E&E &&E&\\
\hline
&&\phantom{E}&&\\
\hline
&&\phantom{E}&&\\
\hline
E&E &&E&\phantom{E}\\
\hline
&E&\phantom{E}&E&\\
\hline
\end{tabular}
&
\begin{tabular}{|c|c|c|c|c|}
\hline
\phantom{E}&E&E& &\\
\hline
&&\phantom{E}&&\\
\hline
&&\phantom{E}&&\\
\hline
&E&E&E&\phantom{E}\\
\hline
&E&E&E&\\
\hline
\end{tabular}
\end{tabular}
}

\vspace{.4cm}

Consider case 1 and assume that the erasures occurred in locations
$(i_0,j_0)$, $(i_0,j_1)$, $(i_0,j_2)$, $(i_0,j_3)$, $(i_1,j_0)$,
$(i_1,j_1)$, $(i_1,j_2)$ and $(i_1,j_3)$, where $0\leq i_0<i_1\leq
m-1$ and $0\leq j_0<j_1<j_2<j_3\leq n-1$.
We need to prove, using the
parity-check matrix as given by~(\ref{pc3}) , that the
$8\times 8$ matrix

$$
\left(
\begin{array}{cccccccc}
1&1&1&1&0&0&0&0\\
0&0&0&0&1&1&1&1\\
0&1&0&0&0&1&0&0\\
0&0&1&0&0&0&1&0\\
0&0&0&1&0&0&0&1\\
\al^{i_0n+j_0}&\al^{i_0n+j_1}&\al^{i_0n+j_2}&\al^{i_0n+j_3}&\al^{i_1n+j_0}&\al^{i_1n+j_1}&\al^{i_1n+j_2}&\al^{i_1n+j_3}\\
\al^{-i_0n-j_0}&\al^{-i_0n-j_1}&\al^{-i_0n-j_2}&\al^{-i_0n-j_3}&\al^{-i_1n-j_0}&\al^{-i_1n-j_1}&\al^{-i_1n-j_2}&\al^{-i_1n-j_3}\\
\al^{2(i_0n+j_0)}&\al^{2(i_0n+j_1)}&\al^{2(i_0n+j_2)}&\al^{2(i_0n+j_3)}&\al^{2(i_1n+j_0)}&\al^{2(i_1n+j_1)}&\al^{2(i_1n+j_2)}&\al^{2(i_1n+j_3)}\\
\end{array}
\right)
$$
is invertible. Redefining $i\la i_1-i_0$, $j_1\la j_1-j_0$,
$j_2\la j_2-j_0$ and $j_3\la j_3-j_0$, where
now $1\leq i\leq m-1$ and $1\leq j_1<j_2<j_3\leq n-1$, this
matrix is invertible if and only if matrix

$$
\left(
\begin{array}{cccccccc}
1&1&1&1&0&0&0&0\\
0&0&0&0&1&1&1&1\\
0&1&0&0&0&1&0&0\\
0&0&1&0&0&0&1&0\\
0&0&0&1&0&0&0&1\\
1&\al^{j_1}&\al^{j_2}&\al^{j_3}&\al^{in}&\al^{in+j_1}&\al^{in+j_2}&\al^{in+j_3}\\
1&\al^{-j_1}&\al^{-j_2}&\al^{-j_3}&\al^{-in}&\al^{-in-j_1}&\al^{-in-j_2}&\al^{-in-j_3}\\
1&\al^{2j_1}&\al^{2j_2}&\al^{2j_3}&\al^{2in}&\al^{2(in+j_1)}&\al^{2(in+j_2)}&\al^{2(in+j_3)}\\
\end{array}
\right)
$$
is invertible. This matrix is invertible if and only if the $7\times
7$ matrix

$$
\left(
\begin{array}{ccccccc}
1&0&0&0&1&0&0\\
1\xor\al^{j_1}&1\xor\al^{j_2}&1\xor\al^{j_3}&\al^{in}&\al^{in+j_1}&\al^{in+j_2}&\al^{in+j_3}\\
1\xor\al^{-j_1}&1\xor\al^{-j_2}&1\xor\al^{-j_3}&\al^{-in}&\al^{-in-j_1}&\al^{-in-j_2}&\al^{-in-j_3}\\
1\xor\al^{2j_1}&1\xor\al^{2j_2}&1\xor\al^{2j_3}&\al^{2in}&\al^{2(in+j_1)}&\al^{(in+j_2)}&\al^{2(in+j_3)}\\
0&1&0&0&0&1&0\\
0&0&1&0&0&0&1\\
0&0&0&1&1&1&1\\
\end{array}
\right)
$$
is invertible, if and only if the $6\times 6$ matrix

$$
\left(
\begin{array}{cccccc}
1&0&0&0&1&0\\
1\xor\al^{j_2}&1\xor\al^{j_3}&\al^{in}&1\xor\al^{j_1}\xor\al^{in+j_1}&\al^{in+j_2}&\al^{in+j_3}\\
1\xor\al^{-j_2}&1\xor\al^{-j_3}&\al^{-in}&1\xor\al^{-j_1}\xor\al^{-in-j_1}&\al^{-in-j_2}&\al^{-in-j_3}\\
1\xor\al^{2j_2}&1\xor\al^{2j_3}&\al^{2in}&1\xor\al^{2j_1}\xor\al^{2(in+j_1)}&\al^{2(in+j_2)}&\al^{2(in+j_3)}\\
0&1&0&0&0&1\\
0&0&1&1&1&1\\
\end{array}
\right)
$$
is invertible, if and only if the $5\times 5$ matrix

$$
\left(
\begin{array}{ccccc}
1&0&0&0&1\\
1\xor\al^{j_3}&\al^{in}&1\xor\al^{j_1}\xor\al^{in+j_1}&1\xor\al^{j_2}\xor\al^{in+j_2}&\al^{in+j_3}\\
1\xor\al^{-j_3}&\al^{-in}&1\xor\al^{-j_1}\xor\al^{-in-j_1}&1\xor\al^{-j_2}\xor\al^{-in-j_2}&\al^{-in-j_3}\\
1\xor\al^{2j_3}&\al^{2in}&1\xor\al^{2j_1}\xor\al^{2(in+j_1)}&1\xor\al^{2j_2}\xor\al^{2(in+j_2)}&\al^{2(in+j_3)}\\
0&1&1&1&1\\
\end{array}
\right)
$$
is invertible, if and only if the $4\times 4$ matrix

$$
\left(
\begin{array}{cccc}
1&1&1&1\\
\al^{in}&1\xor\al^{j_1}\xor\al^{in+j_1}&1\xor\al^{j_2}\xor\al^{in+j_2}&1\xor\al^{j_3}\xor\al^{in+j_3}\\
\al^{-in}&1\xor\al^{-j_1}\xor\al^{-in-j_1}&1\xor\al^{-j_2}\xor\al^{-in-j_2}&1\xor\al^{-j_3}\xor\al^{-in-j_3}\\
\al^{2in}&1\xor\al^{2j_1}\xor\al^{2(in+j_1)}&1\xor\al^{2j_2}\xor\al^{2(in+j_2)}&1\xor\al^{2j_3}\al^{2(in+j_3)}\\
\end{array}
\right)
$$
is invertible, if and only if the $3\times 3$ matrix

$$
\left(
\begin{array}{ccc}
(1\xor\al^{j_1})(1\xor\al^{in})&(1\xor\al^{j_2})(1\xor\al^{in})&(1\xor\al^{j_3})(1\xor\al^{in})\\
(1\xor\al^{-j_1})(1\xor\al^{-in})&(1\xor\al^{-j_2})(1\xor\al^{-in})&(1\xor\al^{-j_3})(1\xor\al^{-in})\\
(1\xor\al^{2j_1})(1\xor\al^{2in})&(1\xor\al^{2j_2})(1\xor\al^{2in})&(1\xor\al^{2j_3})(1\xor\al^{2in})\\
\end{array}
\right)
$$
is invertible, if and only if the $3\times 3$ matrix

$$
\left(
\begin{array}{ccc}
1&1&1\\
\al^{-j_1}&\al^{-j_2}&\al^{-j_3}\\
1\xor\al^{j_1}&1\xor\al^{j_2}&1\xor\al^{j_3}\\
\end{array}
\right)
$$
is invertible, if and only if the $2\times 2$ matrix

$$
\left(
\begin{array}{cc}
\al^{-j_1}\xor\al^{-j_2}&\al^{-j_1}\xor\al^{-j_3}\\
\al^{j_1}\xor\al^{j_2}&\al^{j_1}\xor\al^{j_3}\\
\end{array}
\right)
$$
is invertible, if and only if the $2\times 2$ matrix

$$
\left(
\begin{array}{cc}
1&1\\
\al^{-j_1-j_2}&\al^{-j_1-j_3}\\
\end{array}
\right)
$$
is invertible, if and only if, computing the determinant of the matrix above,
$$\al^{-j_1}(\al^{-j_2}\xor\al^{-j_3})\neq 0,$$
which is true since $1\,\leq\, j_2\,<\,j_3\leq n-1\,<\,\cO(\al)$. So all these cases of 8
erasures are correctable.

Consider next case 2 and assume that the erasures occurred in locations
$(i_0,j_0)$, $(i_0,j_1)$, $(i_1,j_0)$, $(i_1,j_1)$, $(i_2,j_0)$,
$(i_2,j_1)$, $(i_3,j_0)$ and $(i_3,j_1)$, where $0\leq i_0<i_1<i_2<i_3\leq
m-1$ and $0\leq j_0<j_1\leq n-1$.
We need to prove, using the
parity-check matrix as given by~(\ref{pc3}) , that the
$8\times 8$ matrix

$$
\left(
\begin{array}{cccccccc}
1&1&0&0&0&0&0&0\\
0&0&1&1&0&0&0&0\\
0&0&0&0&1&1&0&0\\
0&0&0&0&0&0&1&1\\
0&1&0&1&0&1&0&1\\
\al^{i_0n+j_0}&\al^{i_0n+j_1}&\al^{i_1n+j_0}&\al^{i_1n+j_1}&\al^{i_2n+j_0}&\al^{i_2n+j_1}&\al^{i_3n+j_0}&\al^{i_3n+j_1}\\
\al^{-i_0n-j_0}&\al^{-i_0n-j_1}&\al^{-i_1n-j_0}&\al^{-i_1n-j_1}&\al^{-i_2n-j_0}&\al^{-i_2n-j_1}&\al^{-i_3n-j_0}&\al^{-i_3n-j_1}\\
\al^{2(i_0n+j_0)}&\al^{2(i_0n+j_1)}&\al^{2(i_1n+j_0)}&\al^{2(i_1n+j_1)}&\al^{2(i_2n+j_0)}&\al^{2(i_2n+j_1)}&\al^{2(i_3n+j_0)}&\al^{2(i_3n+j_1)}\\
\end{array}
\right)
$$
is invertible. Redefining $i_1\la i_1-i_0$, $i_2\la i_2-i_0$, $i_3\la
i_3-i_0$ and $j\la j_1-j_0$,
where now $1\leq i_1<i_2<i_3\leq m-1$ and $1\leq j\leq n-1$, this
matrix is invertible if and only if matrix

$$
\left(
\begin{array}{cccccccc}
1&1&0&0&0&0&0&0\\
0&0&1&1&0&0&0&0\\
0&0&0&0&1&1&0&0\\
0&0&0&0&0&0&1&1\\
0&1&0&1&0&1&0&1\\
1&\al^{j}&\al^{i_1n}&\al^{i_1n+j}&\al^{i_2n}&\al^{i_2n+j}&\al^{i_3n}&\al^{i_3n+j}\\
1&\al^{-j}&\al^{-i_1n}&\al^{-i_1n-j}&\al^{-i_2n}&\al^{-i_2n-j}&\al^{-i_3n}&\al^{-i_3n-j}\\
1&\al^{2j}&\al^{2i_1n}&\al^{2(i_1n+j)}&\al^{2i_2n}&\al^{2(i_2n+j)}&\al^{2i_3n}&\al^{2(i_3n+j)}\\
\end{array}
\right)
$$
is invertible. This matrix is invertible if and only if the $4\times
4$ matrix

$$
\left(
\begin{array}{cccc}
1&1&1&1\\
1\xor\al^{j}&\al^{i_1n}(1\xor\al^{j})&\al^{i_2n}(1\xor\al^{j})&\al^{i_3n}(1\xor\al^{j})\\
1\xor\al^{-j}&\al^{-i_1n}(1\xor\al^{-j})&\al^{-i_2n}(1\xor\al^{-j})&\al^{-i_3n}(1\xor\al^{-j})\\
1\xor\al^{2j}&\al^{2i_1n}(1\xor\al^{2j})&\al^{2i_2n}(1\xor\al^{2j})&\al^{2i_3n}(1\xor\al^{2j})\\
\end{array}
\right)
$$
is invertible, if and only if the $3\times 3$ matrix

$$
\left(
\begin{array}{ccc}
(1\xor\al^{i_1n})(1\xor\al^{j})&(1\xor\al^{i_2n})(1\xor\al^{j})&(1\xor\al^{i_3n})(1\xor\al^{j})\\
(1\xor\al^{-i_1n})(1\xor\al^{-j})&(1\xor\al^{-i_2n})(1\xor\al^{-j})&(1\xor\al^{-i_3n})(1\xor\al^{-j})\\
(1\xor\al^{2i_1n})(1\xor\al^{2j})&(1\xor\al^{2i_2n})(1\xor\al^{2j})&(1\xor\al^{2i_3n})(1\xor\al^{2j})\\
\end{array}
\right)
$$
is invertible, if and only if the $3\times 3$ matrix

$$
\left(
\begin{array}{ccc}
1&1&1\\
\al^{-i_1n}&\al^{-i_2n}&\al^{-i_3n}\\
1\xor\al^{i_1n}&1\xor\al^{i_2n}&1\xor\al^{i_3n}\\
\end{array}
\right)
$$
is invertible, if and only if the $2\times 2$ matrix

$$
\left(
\begin{array}{cc}
\al^{i_1n}\xor\al^{i_2n}&\al^{i_1n}\xor\al^{i_3n}\\
\al^{-i_1n}\xor\al^{-i_2n}&\al^{-i_1n}\xor\al^{-i_3n}\\
\end{array}
\right)\;\eq\;
(\al^{i_1n}\xor\al^{i_2n})(\al^{i_1n}\xor\al^{i_3n})
\left(
\begin{array}{cc}
1&1\\
\al^{(-i_1-i_2)n}&\al^{(-i_1-i_2)n}\\
\end{array}
\right)
$$
is invertible, if and only if, computing the determinant of this last matrix,

$$\al^{(-i_1-i_2)n}\xor\al^{(-i_1-i_2)n}\eq\al^{-i_1n}(\al^{-i_2n}\xor
\al^{-i_3n})\neq 0,$$
which is certainly the case since $1\leq i_2n\,<\,i_3n\,<\,mn\,<\,\cO(\al)$.

Consider finally case 3.
Let $i_0<i_1$ and $j_0<j_1$ be the rows and columns respectively with
three erasures.
It suffices to prove, using the
parity-check matrix $\cH_3$ given by~(\ref{pc3}), that the
$8\times 8$ matrix

$$
\left(
\begin{array}{cccccccc}
1&1&1&0&0&0&0&0\\
0&0&0&1&1&1&0&0\\
0&0&0&0&0&0&1&1\\
0&1&0&0&1&0&0&1\\
0&0&1&0&0&1&0&0\\
\al^{i_0n+j_0}&\al^{i_0n+j_1}&\al^{i_0n+j_2}&\al^{i_1n+j_0}&\al^{i_1n+j_1}&\al^{i_1n+j_2}&\al^{i_2n+j_0}&\al^{i_2n+j_1}\\
\al^{-i_0n-j_0}&\al^{-i_0n-j_1}&\al^{-i_0n-j_2}&\al^{-i_1n-j_0}&\al^{-i_1n-j_1}&\al^{-i_1n-j_2}&\al^{-i_2n-j_0}&\al^{-i_2n-j_1}\\
\al^{2(i_0n+j_0)}&\al^{2(i_0n+j_1)}&\al^{2(i_0n+j_2)}&\al^{2(i_1n+j_0)}&\al^{2(i_1n+j_1)}&\al^{2(i_1n+j_2)}&\al^{2(i_2n+j_0)}&\al^{2(i_2n+j_1)}\\
\end{array}
\right)
$$
is invertible.

Redefining $i_1\la i_1-i_0$, $i_2\la i_2-i_0$, $j_1\la j_1-j_0$ and $j_2\la j_2-j_0$,
the matrix above is invertible if and only if the matrix

$$
\left(
\begin{array}{cccccccc}
1&1&1&0&0&0&0&0\\
0&0&0&1&1&1&0&0\\
0&0&0&0&0&0&1&1\\
0&1&0&0&1&0&0&1\\
0&0&1&0&0&1&0&0\\
1&\al^{j_1}&\al^{j_2}&\al^{i_1n}&\al^{i_1n+j_1}&\al^{i_1n+j_2}&\al^{i_2n}&\al^{i_2n+j_1}\\
1&\al^{-j_1}&\al^{-j_2}&\al^{-i_1n-j_0}&\al^{-i_1n-j_1}&\al^{-i_1n-j_2}&\al^{-i_2n}&\al^{-i_2n-j_1}\\
1&\al^{2j_1}&\al^{2j_2}&\al^{2i_1n}&\al^{2(i_1n+j_1)}&\al^{2(i_1n+j_2)}&\al^{2i_2n}&\al^{2(i_2n+j_1)}\\
\end{array}
\right)
$$
is invertible, if and only if the matrix

$$
\left(
\begin{array}{ccccccc}
1&0&0&1&0&0&1\\
1\xor\al^{j_1}&1\xor\al^{j_2}&\al^{i_1n}&\al^{i_1n+j_1}&\al^{i_1n+j_2}&\al^{i_2n}&\al^{i_2n+j_1}\\
1\xor\al^{-j_1}&1\xor\al^{-j_2}&\al^{-i_1n-j_0}&\al^{-i_1n-j_1}&\al^{-i_1n-j_2}&\al^{-i_2n}&\al^{-i_2n-j_1}\\
1\xor\al^{2j_1}&1\xor\al^{2j_2}&\al^{2i_1n}&\al^{2(i_1n+j_1)}&\al^{2(i_1n+j_2)}&\al^{2i_2n}&\al^{2(i_2n+j_1)}\\
0&1&0&0&1&0&0\\
0&0&1&1&1&0&0\\
0&0&0&0&0&1&1\\
\end{array}
\right)
$$

is invertible, if and only if the matrix

$$
\left(
\begin{array}{cccccc}
1&0&0&1&0&0\\
1\xor\al^{j_2}&\al^{i_1n}&1\xor\al^{j_1}\xor\al^{i_1n+j_1}&\al^{i_1n+j_2}    &\al^{i_2n}&1\xor\al^{j_1}\xor\al^{i_2n+j_1}\\
1\xor\al^{-j_2}&\al^{-i_1n}&1\xor\al^{-j_1}\xor\al^{-i_1n-j_1}&\al^{-i_1n-j_2}&\al^{-i_2n}&1\xor\al^{-j_1}\xor\al^{-i_2n-j_1}\\
1\xor\al^{2j_2}&\al^{2i_1n}&1\xor\al^{2j_1}\xor\al^{2(i_1n+j_1)}&\al^{2(i_1n+j_2)}&\al^{2i_2n}&1\xor\al^{2j_1}\xor\al^{2(i_2n+j_1)}\\
0&1&1&1&0&0\\
0&0&0&0&1&1\\
\end{array}
\right)
$$

is invertible, if and only if the matrix

$$
\left(
\begin{array}{ccccc}
1&1&1&0&0\\
\al^{i_1n}&1\xor\al^{j_1}\xor\al^{i_1n+j_1}&1\xor\al^{j_2}\xor\al^{i_1n+j_2}&\al^{i_2n}&1\xor\al^{j_1}\xor\al^{i_2n+j_1}\\
\al^{-i_1n}&1\xor\al^{-j_1}\xor\al^{-i_1n-j_1}&1\xor\al^{-j_2}\xor\al^{-i_1n-j_2}&\al^{-i_2n}&1\xor\al^{-j_1}\xor\al^{-i_2n-j_1}\\
\al^{2i_1n}&1\xor\al^{2j_1}\xor\al^{2(i_1n+j_1)}&\al^{2(i_1n+j_2)}&1\xor\al^{2j_1}\xor\al^{2i_2n}&1\xor\al^{2j_2}\xor
\al^{2(i_2n+j_1)}\\
0&0&0&1&1\\
\end{array}
\right)
$$

is invertible, if and only if the matrix

$$
\left(
\begin{array}{cccc}
(1\xor\al^{i_1n})(1\xor\al^{j_1})&(1\xor\al^{i_1n})(1\xor\al^{j_2})&\al^{i_2n}&1\xor\al^{j_1}\xor\al^{i_2n+j_1}\\
(1\xor\al^{-i_1n})(1\xor\al^{-j_1})&(1\xor\al^{-i_1n})(1\xor\al^{-j_2})&\al^{-i_2n}&1\xor\al^{-j_1}\xor\al^{-i_2n-j_1}\\
(1\xor\al^{2i_1n})(1\xor\al^{2j_1})&(1\xor\al^{2i_1n})(1\xor\al^{2j_2})&\al^{2i_2n}&1\xor\al^{2j_1}\xor\al^{2(i_2n+j_1)}\\
0&0&1&1\\
\end{array}
\right)
$$

is invertible, if and only if the matrix

$$
\left(
\begin{array}{ccc}
(1\xor\al^{i_1n})(1\xor\al^{j_1})&(1\xor\al^{i_1n})(1\xor\al^{j_2})&(1\xor\al^{i_2n})(1\xor\al^{j_1})\\
(1\xor\al^{-i_1n})(1\xor\al^{-j_1})&(1\xor\al^{-i_1n})(1\xor\al^{-j_2})&(1\xor\al^{-i_2n})(1\xor\al^{-j_1})\\
(1\xor\al^{2i_1n})(1\xor\al^{2j_1})&(1\xor\al^{2i_1n})(1\xor\al^{2j_2})&(1\xor\al^{2i_2n})(1\xor\al^{2j_1})\\
\end{array}
\right)
$$

is invertible, if and only if the matrix

$$
\left(
\begin{array}{ccc}
1&1&1\\
\al^{-i_1n-j_1}&\al^{-i_1n-j_2}&\al^{-i_2n-j_1}\\
(1\xor\al^{i_1n})(1\xor\al^{j_1})&(1\xor\al^{i_1n})(1\xor\al^{j_2})&(1\xor\al^{i_2n})(1\xor\al^{j_1})\\
\end{array}
\right)
$$

is invertible, if and only if the matrix

$$
\left(
\begin{array}{cc}
\al^{-i_1n-j_1}(\al^{-j_1}\xor\al^{-j_2})&\al^{-j_1}(\al^{-i_1n}\xor\al^{-i_2n})\\
(1\xor\al^{i_1n})(\al^{j_1}\xor\al^{j_2})&(1\xor\al^{j_1})(\al^{i_1n}\xor\al^{i_2n})\\
\end{array}
\right)
$$

is invertible, if and only if the matrix

$$
\left(
\begin{array}{cc}
\al^{-i_1n-2j_1-j_2}&\al^{-(i_1+i_2)n-j_1}\\
1\xor\al^{i_1n} &1\xor\al^{j_1}\\
\end{array}
\right)
$$

is invertible, if and only if the matrix

$$
\left(
\begin{array}{cc}
\al^{-j_1-j_2}&\al^{-i_2n}\\
1\xor\al^{i_1n} &1\xor\al^{j_1}\\
\end{array}
\right)
$$
which is invertible
if and only if its determinant, which is given by the left hand side
of~(\ref{inv1}) times a constant, is non-zero.

\qed

Notice that in Lemma~\ref{lemma5}, 8 erasures following the patterns
of cases~1 and~2 will always be corrected, while case~3 will be
corrected only when condition~(\ref{inv1}) is satisfied. So Lemma~\ref{lemma5}
by itself does not prove that there is an
$EP(m,1;n,1;3)$ code with minimum distance 9, but we can find a code
satisfying~(\ref{inv1}) using an argument similar to the one used to
show an infinite family of PMDS codes in~\cite{bhh}. In effect,
consider the field $GF(2^p)$, $p$ a prime number, such that $GF(2^p)$
is generated by the irreducible polynomial $M_p(x)\eq 1+x+x^2+\cdots
+x^{p-1}$. The polynomial $M_p(x)$ may not be irreducible, for
example, $M_5(x)$ is irreducible but $M_7(x)\eq
(1+x+x^3)(1+x^2+x^3)$, so not any prime number can be chosen. If we
choose a prime number large enough, condition~(\ref{inv1}) will
hold, as shown in the next corollary:

\begin{cor}
\label{cor5}
{\em
Consider the $EP(m,1;n,1;3)$ code whose parity-check matrix is given
by~(\ref{pc3}) with $\al$ in~(\ref{pc3}) a zero of $M_p(x)$, $p$ a
prime number, $M_p(x)$ irreducible and
$mn<p$. Then the code has minimum distance 9.
}
\end{cor}

\noindent\pf We have to show that~(\ref{inv1}) is satisfied.
Given an integer $z$, denote by $\lan z\ran_p$ the unique integer
$u$, $0\leq u\leq p-1$, such that $u\equiv z\;\;(\bmod\;\;p)$. Let
$M_p(\al)\eq 0$, then $\cO(\al)\eq p$. Hence,

\begin{eqnarray*}
1\xor\al^{-j_1}\xor\al^{-i_2n+j_2}\xor\al^{-(i_2-i_1)n+j_2}&\eq &
1\xor\al^{\lan -j_1\ran_p}\xor\al^{\lan -i_2n+j_2\ran_p}\xor\al^{\lan
-(i_2-i_1)n+j_2\ran_p}.
\end{eqnarray*}

Take the
first three elements in~(\ref{inv1}), i.e.,

\begin{eqnarray*}
1\xor\al^{-j_1}\xor\al^{-i_2n+j_2}\eq 1\xor\al^{\lan -j_1\ran_p}\xor\al^{\lan -i_2n+j_2\ran_p}.
\end{eqnarray*}

Since $1\leq j_1\leq n-1$, $1\leq |j_2|\leq n-1$, $1\leq |i_2|\leq
m-1$ and $mn<p$, $\lan -j_1\ran_p\eq p-j_1\neq 0$.

Assume that $\lan -i_2n+j_2\ran_p\eq 0$, then $-i_2n+j_2\eq sp$ for
some integer $s$. If $s\eq 0$, then $i_2n\eq j_2$, a contradiction
since $1\leq |j_2|\leq n-1$ and $1\leq |i_2|\leq m-1$. So $s\neq 0$
and $j_2\eq sp+i_2n$. If $s\geq 1$, since $1\leq |i_2|\leq m-1$ and $mn<p$,
$j_2\eq sp+i_2n\geq mn-(m-1)n\eq n$, a contradiction since $1\leq
|j_2|\leq n-1$. If $s<0$, $j_2\eq sp+i_2n\leq -mn+(m-1)n\eq -n$, also
a contradiction.

If $\lan -j_1\ran_p\neq\lan -i_2n+j_2\ran_p$, then the
first three elements in~(\ref{inv1}) are distinct from each other, so
they cannot be canceled by the 4th element and~(\ref{inv1}) holds.
So, assume that $\lan -j_1\ran_p\eq\lan -i_2n+j_2\ran_p$. In particular,

\begin{eqnarray}
\label{i2n}
-i_2n+j_2&=&-j_1+sp\;\;\;{\rm for}\;\;{\rm some}\;\;{\rm integer}\;\;s.
\end{eqnarray}

Now, in order for the left hand side of~(\ref{inv1}) to be zero, in
addition to~(\ref{i2n}), we need $\lan -(i_2-i_1)n+j_2\ran_p\eq 0$, giving

\begin{eqnarray}
\label{i2ni1n}
-i_2n+j_2&=&-i_1n+s'p\;\;\;{\rm for}\;\;{\rm some}\;\;{\rm integer}\;\;s'.
\end{eqnarray}

Combining~(\ref{i2n}) and~(\ref{i2ni1n}), we obtain

\begin{eqnarray}
\label{i1n}
i_1n-j_1&=&s''p\;\;\;{\rm for}\;\;{\rm some}\;\;{\rm integer}\;\;s''.
\end{eqnarray}

Since $1\leq i_1\leq m-1$ and $1\leq j_1\leq n-1$,

$$
1\leq i_1n-j_1\,<\, mn<p,
$$
contradicting~(\ref{i1n}) and completing the proof.

\qed

Corollary~\ref{cor5} shows that bound~(\ref{d3s}) is indeed tight.

The construction in Corollary~\ref{cor5} depends on, for each
$m\times n$ array, finding a prime $p$ such that $mn\,<\,p$ and $M_p(x)$ is
irreducible (it is well known that $M_p(x)$ is irreducible if and
only if 2 is primitive in $GF(p)$~\cite{ms}). Strictly speaking, it
is not proven that the number of such primes is infinite, but it is
believed it is, and from a practical point of view, it is always
possible to find such a large enough prime number.

Let us point out that although the field of polynomials modulo
$M_p(x)$ has size $2^p$, no look-up tables are necessary in
implementation, since most operations reduce to XORs and
rotations~\cite{br}. We omit the details here.

\section{Conclusions}
\label{conclusions}
We have studied extended product (EPC) codes, in which a few global
parities are added to a traditional product code in order to enhance
its distance properties. We presented a special case of extended
product codes: generalized product (GPC) codes. We showed that GPC
codes unify two types of codes: product codes and integrated
interleaved (II) codes. We studied the distance properties of these
type of codes. Although, except for the special case of one global
parity, GPC codes do not optimize the minimum distance, they can be
implemented with modest field size, and in addition they provide a
large variety of possible parameters, making them an attractive
alternative for implementation in practical cases. We showed some
optimal constructions for two and three global parities 
requiring a larger field size.

\end{document}